\documentclass{article}

\usepackage[nonatbib,final]{neurips_data_2024}

\usepackage[numbers]{natbib}
\usepackage[utf8]{inputenc} 
\usepackage[T1]{fontenc}    
\usepackage{hyperref}       
\usepackage{url}            
\usepackage{booktabs}       
\usepackage{amsfonts}       
\usepackage{nicefrac}       
\usepackage{microtype}      
\usepackage{xcolor}         
\usepackage{microtype}
\usepackage{graphicx}
\usepackage{subfigure}
\usepackage{booktabs} 
\usepackage[normalem]{ulem}
\usepackage{adjustbox}
\usepackage{amsmath}
\usepackage{amssymb}
\usepackage{mathtools}
\usepackage{array}
\usepackage{enumitem}
\usepackage{subfigure}
\usepackage{colortbl}
\usepackage{soul}
\usepackage{multirow}
\usepackage{makecell}
\definecolor{mypink}{rgb}{.99,.90,.92}
\definecolor{magicmint}{rgb}{0.67, 0.94, 0.82}
\usepackage{tcolorbox}

\usepackage{listings}
\usepackage{pythonhighlight}

\usepackage{titletoc}
\usepackage{letltxmacro}
\usepackage{etoolbox}
\usepackage{lipsum}
\newcommand\MiniToC{%
  \setcounter{tocdepth}{2}
  \startcontents
  \printcontents{}{1}{\section*{\contentsname}\vskip-3.5ex\hrulefill\vskip1ex}
  \vskip-0.5ex\noindent\hrulefill
  \setcounter{tocdepth}{0}
}
\definecolor{mypink}{rgb}{.99,.91,.95}
\definecolor{mygrey}{rgb}{.9,.9,.9}

\usepackage{soul,color,xcolor}
\usepackage[capitalize,noabbrev]{cleveref}
\colorlet{soulblue}{blue!20}
\colorlet{yw}{orange!50}
\definecolor{mypurple}{rgb}{.80, .70, .84}
\definecolor{myblue}{rgb}{.86, .91, .95}

\newcommand{\reb}[1]{\textcolor{black}{#1}}

\title{Towards Open Respiratory Acoustic Foundation Models: Pretraining and Benchmarking }

\author{%
  Yuwei Zhang$\dag^1$, 
   Tong Xia$\dag^1$, 
   Jing Han$^1$, 
  Yu Yvonne Wu$^1$, 
   Georgios Rizos$^1$, 
   Yang Liu$^1$,  \\
 \textbf{Mohammed Mosuily$^2$}, 
 \textbf{Jagmohan Chauhan$^2$}, 
  \textbf{Cecilia Mascolo$^1$} \\
 $^1$ University of Cambridge, $^2$ University of Southampton, UK\\
$\dag$ joint first authors, equal contribution\\
  \texttt{\{yz798, tx229\}@cam.ac.uk} \\
}

\begin{document}

\maketitle

\vspace{-10pt}
\begin{abstract}
Respiratory audio, such as coughing and breathing sounds, has predictive power for a wide range of healthcare applications, yet is currently under-explored. The main problem for those applications arises from the difficulty in collecting large labeled task-specific data for model development. Generalizable respiratory acoustic foundation models pretrained with unlabeled data would offer appealing advantages and possibly unlock this impasse.  However, given the safety-critical nature of healthcare applications, it is pivotal to also ensure openness and replicability for any proposed foundation model solution. To this end, we introduce OPERA, an OPEn Respiratory Acoustic foundation model pretraining and benchmarking system, as the first approach answering this need. We curate large-scale respiratory audio datasets ($\sim$136K samples, \reb{over 400} hours), pretrain three pioneering \reb{generalizable acoustic models}, and build a benchmark consisting of 19 downstream respiratory health tasks for evaluation. Our pretrained models demonstrate superior performance (against existing acoustic models pretrained with general audio on 16 out of 19 tasks) and generalizability (to unseen datasets and new respiratory audio modalities). This highlights the great promise of respiratory acoustic foundation models and encourages more studies using OPERA as an open resource to accelerate research on respiratory audio for health. 
\end{abstract}

{%
    \hypersetup{
        colorlinks=true,
        urlcolor=blue, 
        linkcolor=blue, 
        citecolor=blue  
    }
\vspace{-7pt}
\begin{adjustbox}{width=320pt,center}
\begin{tcolorbox}[colframe=black, colback=gray!8, boxrule=0.8pt, width=0.9\textwidth, arc=2mm, left=2mm, right=2mm, top=0mm, bottom=0mm, boxsep=2mm]

    $ \ \ \ \ \ \ \ \ \ \ \ $ The OPERA website can be found at \href{https://opera-benchmark.github.io}{opera-benchmark.github.io} 

    \vspace{3pt}
    $ \ \ \ \ \ \ \ \ \ \ \ $ Our codebase is open-sourced at \href{https://github.com/evelyn0414/OPERA}{github.com/evelyn0414/OPERA}
\end{tcolorbox}
\end{adjustbox}
}


\section{Introduction}

Respiratory audio, such as coughing and breathing sounds generated by the respiratory system's airflow, contains multiple physiological characteristics of individuals and therefore its modeling could be instrumental in health monitoring and disease detection applications~\cite{reichert2008analysis, sovijarvi2000definition}. For instance, audio recordings can be used to estimate respiratory rate and lung function~\cite{doheny2023estimation, rudraraju2020cough, xie2023earspiro}, detect snoring and apnea events during sleep~\cite{jane2000automatic, halevi2016can, romero2022acoustic}, assess the effect of smoking on health~\cite{ma2022determining, ma2023towards} and diagnose diseases like flu and asthma~\cite{jayadi2022embedded, islam2018multichannel,han2022sounds,rizos2023positive}.

To enable the widespread adoption of these applications, high-performing algorithms are needed.
Related studies rely on traditional signal processing methods~\cite{doheny2023estimation, rudraraju2020cough, jane2000automatic,halevi2016can, ma2022determining, jayadi2022embedded, islam2018multichannel}, which require domain knowledge and often exhibit limited performance. Supervised deep acoustic models have been proposed~\cite{xie2023earspiro, han2022sounds,srivastava2021deep} 
but their performance heavily depends on the volume and quality of available labels, which might be difficult and expensive to collect. Hence, foundation models pretrained with large unlabeled respiratory audio data have a high potential to improve performance 
through transfer learning and supervised fine-tuning~\cite{dang2023human, xia2022exploring}. However, in contrast with other health data modalities like clinical imaging~\cite{pai2024foundation}, electronic health records (EHRs)~\cite{wornow2023shaky}, and medical time series~\cite{yeh2023toward,abbaspourazad2023large,dong2024simmtm}, foundation models for respiratory audio are largely under-explored. 

\textbf{Respiratory audio datasets are available but no comprehensive collection has been curated.}
Recent years have seen an ever-increasing accumulation of respiratory audio~\cite{xia2021covid, orlandic2021coughvid, coppock2024audio}, exhibiting heterogeneous properties such as varying acquisition
modalities and sampling rates. These datasets exhibit significant potential for acoustic model development and evaluation. However, no existing effort has curated such data systematically.

\textbf{There is no open respiratory acoustic foundation model, 
impeding the field's growth and understanding.}
Existing open-source acoustic models like AudioMAE~\cite{huang2022masked} and CLAP~\cite{elizalde2023clap} are pretrained on general audio event datasets such as YouTube audio, containing very few (around 0.3\%) respiratory sounds~\cite{jansen2017large, gemmeke2017audio}. These models may not be able to effectively capture the subtle nuances of respiratory sounds, which can vary in abrupt bursts, aperiodic components, and frequency distributions, particularly across different health conditions~\cite{reichert2008analysis}. 
Although a model pretrained on respiratory sounds has been recently presented~\cite{baur2024hear}, it is not open-source, making it hard to analyze, replicate, or compare its workings.
The insights on how to effectively train \reb{generalizable} respiratory acoustic models also remain limited.

\textbf{There is no ready-to-use benchmark for respiratory audio research.}
Current task-specific studies evaluate their models on purposely collected datasets, 
leaving the models' generalizability to other tasks unclear~\cite{baur2024hear}. A benchmark that combines multiple public datasets across diverse applications to enable fair and comprehensive evaluations of the developed foundation models is essential but currently lacking.  This is crucial for safety-critical health applications, where models must be rigorously evaluated before use~\cite{wu2024towards,vollmer2020machine}.

\begin{figure}[t]
\begin{adjustbox}{width=0.99\textwidth,center}
    \centering
    \includegraphics[width=\textwidth]{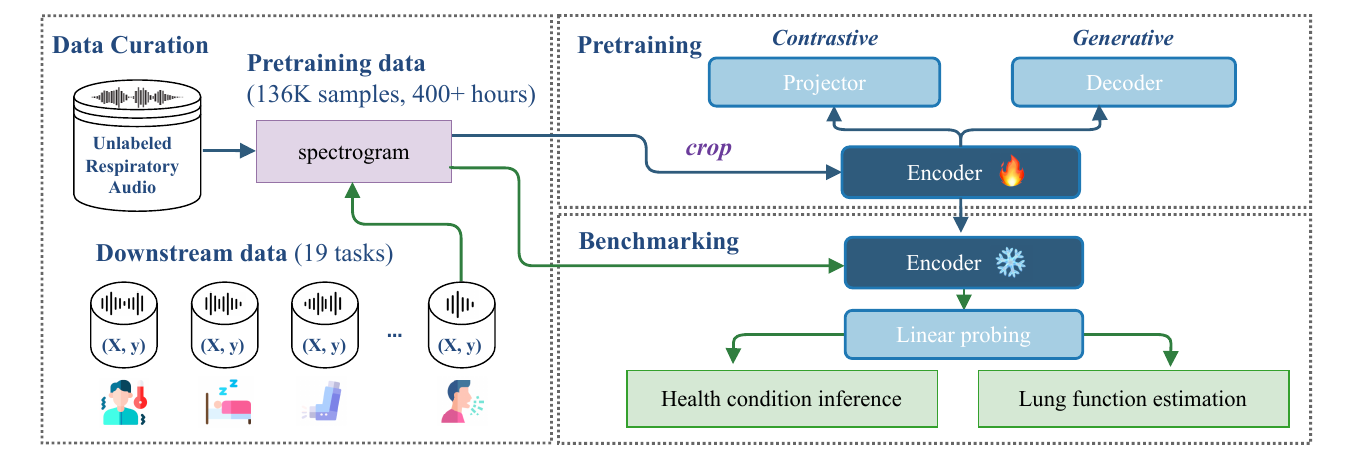}
    \end{adjustbox}
    \caption{System overview of OPERA. After data curation, respiratory audio \reb{encoders} are pretrained and then evaluated on various downstream health tasks.
    }
    \label{fig:framework}
    \vspace{-10pt}
\end{figure}


To mitigate these gaps, in this paper, we put forward \textit{OPERA}, an \textbf{OPE}n \textbf{R}espiratory \textbf{A}coustic foundation model pretraining and benchmarking system (\cref{fig:framework}). It curates unlabeled respiratory audio datasets, pretrain three pioneering \reb{foundational models}, and evaluates them against existing pretrained acoustic models across various applications.  Specifically, our contributions are:
\begin{itemize}[topsep=-1pt, itemsep=-1pt, leftmargin=15pt]
\item We curate a unique large-scale ($\sim$136K samples, \reb{400+} hours), multi-source (5 datasets), multi-modal (breathing, coughing, and lung sounds) and publicly available (or available on request) respiratory audio dataset for \reb{generalizable} model pretraining, 
orders of magnitude larger than the number of respiratory audio samples in datasets used for training existing open acoustic models. 

\item We pretrain 3 \reb{generalizable acoustic} models with the curated unlabeled data using the most common self-supervised  approaches (a contrastive learning-based transformer, a contrastive learning-based CNN model, and a generatively pretrained transformer) to study the effect of the training designs.

\item We employ 10 labeled datasets (6 not covered by pretraining) to formulate 19 respiratory health tasks (12 in health condition inference and 7 in lung function estimation),  ensuring fair, comprehensive and reproducible downstream evaluation. 
\item We benchmark the performance of our 3 \reb{pretrained} models, one commonly used acoustic feature set, and 3 open pretrained acoustic models on these tasks as a starting point for future exploration. 


\end{itemize}

Extensive experiments demonstrate that our \reb{pretrained} models outperform the models pretrained with general audio on 16 out of 19 benchmark tasks, confirming the power and promise of dedicated respiratory acoustic foundation models. Results also show that our models are generalizable across multiple downstream tasks, including new datasets and unseen respiratory audio modalities. 
This is a critical advancement towards realizing the potential of respiratory sounds as a mainstream technique for health monitoring.

Within our three models, we find that the contrastive pretraining model is better for classification-based downstream tasks, while the generative pretrained model performs better in regression tasks, possibly due to the nature of their training objectives: contrastive learning can capture the nuances of the
local patterns to make features distinguishable while generative learning focuses more on global features which are vital for regression.
Our transformer models generally outperform the CNN model
because they have stronger modeling capability, though requiring more intensive computation.
These findings provide insightful guidance to the development and application of such types of models. 
In summary, this paper introduces \textit{the first open-source respiratory acoustic foundation model pretraining and benchmarking system}. This represents a critical first step towards 
comprehensive and reproducible audio foundation models for health: future foundation model research can leverage our system as an experimental resource, and application studies can take advantage of our \reb{pretrained} models as feature extractors. This can facilitate progress in both machine learning and  healthcare. 
These efforts will extend current machine learning capabilities, now able to \textit{see} (via vision) and \textit{read} (via natural languages), to also \textit{listen to} (via audio) our health.

\vspace{-5pt}
\section{Related Work}

\vspace{-5pt}
\subsection{Pretraining in Acoustic Modeling}\label{sec:relate_model}
\vspace{-3pt}

 


Models pretrained on large-scale datasets 
have demonstrated great generalizability in diverse downstream tasks, especially when labeled data are limited~\cite{brown2020language, dosovitskiy2020image, gong2021ast, huang2022masked}. For audio-driven health applications, several general audio pretrained models can be used as feature extractors. One widely used model is \textit{VGGish}~\cite{hershey2017cnn}, trained on 5.24 million hours of audio from YouTube videos to predict 30,871 categories of video labels. Other models have been developed for audio event classification tasks~\cite{kong2020panns, chen2022hts, huang2022masked}. Among them, \textit{AudioMAE}~\cite{huang2022masked} is an open model trained via an auto-encoding objective without requiring any audio labels. Inspired by recent advances in large language models, language-supervised pretraining has also been explored. \textit{CLAP}~\cite{elizalde2023clap} is an open model pretrained in this manner. 
We have included these open models in our benchmark.

It is also worth noting that these open models are pretrained on general audio event datasets such as \textit{AudioSet}~\cite{gemmeke2017audio}, \textit{FSD50K}~\cite{fonseca2021fsd50k}, and \textit{FreeSound}~\cite{fonseca2017freesound}, which contains few samples of respiratory-related audio. For instance, AudioSet's 2 million clips include only 2334 snoring,   871 cough, 834 breathing, and 1200 sneeze clips, making up only 0.3\% of the total. 
In face of this issue, we curate large-scale respiratory audio datasets to pretrain our \reb{generalizable respiratory acoustic} models for comparison. 

In terms of pretraining methods, given the difficulty in collecting large-scale labeled health-related datasets, we consider self-supervised learning (SSL) to leverage unlabelled data for learning meaningful representations~\cite{tang2021selfhar,abbaspourazad2023large, baur2024hear}.  Main SSL methods fall into two categories: contrastive~\cite{chen2020simple, baevski2020wav2vec, saeed2021contrastive} and generative~\cite{he2022masked, huang2022masked, niizumi2023masked}. Contrastive learning trains models to distinguish between similar and dissimilar samples, while generative models are trained to reconstruct original audio data or features from masked or corrupted versions. Since they have been demonstrated to be effective in general audio, We implement both methods in our system.

A recent work, \textit{HeAR}~\cite{baur2024hear}, curated millions of respiratory audio clips from YouTube videos to pretrain a model using a generative SSL approach. However, neither the data nor the model are publicly available, resulting in a lack of transparency and reproducibility. Limited exploration has been conducted on the reasoning behind the chosen SSL method for various downstream tasks.
Our work investigates, for the first time, open pretraining \reb{generalizable} respiratory acoustic models to provide a better understanding of their limits and their potential.


\vspace{-5pt}
\subsection{Benchmarks in Respiratory Audio-based Applications}
\vspace{-5pt}
Current respiratory audio-based health studies typically evaluate their developed models using their self-formulated protocols~\cite{ baur2024hear, xiao2023snoring, xue2021exploring}, instead of following a uniform evaluation pipeline.
This leads to weak reproducibility due to several challenges~\cite{han2022sounds}: lack of implementation details or released code, absence of reliable training and testing division, and varying implementation frameworks (e.g., some in TensorFlow~\cite{han2022sounds} while other in PyTorch~\cite{bae2023patch}) making them difficult to compare.

High-quality benchmarks are essential in machine learning to ensure advancements are reliable and applicable to real-world problems.  While several benchmarks exist for pretrained representation models on general audio event detection and speech recognition~\cite{turian2022hear, sharan2021benchmarking, grollmisch2021analyzing, yang2024air}, similar benchmarks are missing in respiratory audio for health, despite their equal importance. 
The only related benchmark~\cite{hsu2021benchmarking} in this area compares supervised models for breath phase and adventitious sound detection using a single dataset, and is thus not applicable for evaluating foundation models. A comprehensive benchmarking effort of respiratory acoustic foundation models is lacking but has the potential to really shed light on the power of these techniques in the context of respiratory health tasks.


\vspace{-5pt}
\section{System Overview}
\vspace{-5pt}
As shown in \cref{fig:framework}, OPERA comprises three main components: data curation (including unlabeled data for pretraining and labeled data for evaluation), general-purpose pretraining to develop \reb{generalizable} acoustic models (Encoder), and a benchmark comparing the pretrained models on various downstream tasks.

In OPERA, we employ five datasets for pretraining and ten datasets for benchmarking. Four of the downstream datasets overlap with the pretraining resources, but we ensure the testing data is held out before pretraining and thus is never seen by the models.  During the pretraining step, we build two SSL strategies enabling the use of different encoder architectures. We then use the pretrained models to extract features and apply linear probing to report the performance for downstream tasks. Detailed information about data curation and pretraining methods is elaborated on in Section~\ref{sec:pretraining}, and the benchmark data curation and evaluation results are summarized in Section~\ref{sec:benchmark}. 


\section{Self-supervised Pretraining}\label{sec:pretraining}
\vspace{-5pt}
\subsection{Pretraining Datasets}\label{sec:pretraining data}
\vspace{-5pt}
Five open data resources are curated in OPERA to enable the training of respiratory acoustic foundation models (\cref{tab:data}). 
They were collected by different research institutions using various protocols, and are all publicly available or accessible upon request. Some recordings were made with a microphone near the mouth~\cite{xia2021covid, coppock2024audio,orlandic2021coughvid}, while others used a digital stethoscope attached to the chest~\cite{rocha2019open, hsu2022progressively}. This allows the pretrained models to see heterogeneous data for better generalizability.

We only include qualified samples (those identified as respiratory audio, not noise) in the pretraining step. Some labeled audio samples from these datasets, which can be used for downstream evaluations, are held out. We then trim the remaining audio recordings by removing the beginning and ending silence to further ensure the quality of the data. The statistics of the data after quality check are summarized in \cref{tab:data} (extended description can be found in \cref{app:dataset}). As a result, the entire pretraining dataset consists of 135,944 samples, with a total duration of about 404.1 hours.

Before pretraining, all recordings are resampled to 16~kHz and merged into a mono channel. They are then transformed into spectrograms using 64 Mel filter banks with a 64 ms Hann window that shifts every 32 ms~\cite{shen2018natural, zhou2021cough}. For example, a 4s recording will be converted into a spectrogram of $1 \times 126 \times 64$ dimension. 
Finally, these spectrograms are used to pretrain our respiratory acoustic foundation models.

\begin{table}[t]
\caption{Statistics of the data used for model pretraining (SR: sampling rate; Duration: mean [95\% quantile range]; Crop: cropped length for pretraining). }
\centering
\resizebox{0.99\textwidth}{!}
{
\begin{tabular}{llllllc}
\toprule
\textbf{Data name}  &\textbf{Collected by} & \textbf{SR} & \textbf{Modality}  &\#\textbf{Sample}   &\textbf{Duration (s)}  &\textbf{Crop (s)}\\
  \midrule
COVID-19~Sounds~\cite{xia2021covid} &Microphone & 16$\sim$44.1kHz    &Induced cough (3 times)&40866    & 6.1[2.6$\sim$11.2] & 2\\
                                    &                                  &&Deep breath (5 times)&36605   & 20.5[9.7$\sim$31.6] &8\\  
UK~COVID-19~\cite{coppock2024audio}   &Microphone  &48kHz               &Induced cough (3 times) & 19533  & 4.1[2.1$\sim$9.2] & 2\\
                                                                    &   &&Exhalation (5 times) & 20719  &7.7[4.2$\sim$15.6] & 4\\  
COUGHVID~\cite{orlandic2021coughvid} &Microphone  &48kHz             &Induced cough (up to 10s)& 7179   & 6.9[2.4$\sim$9.9] &2\\  

ICBHI~\cite{rocha2019open} &Stethoscope  &4$\sim$44.1kHz             &lung sound (several breath cycles) & 538  & 22.2[20.0$\sim$65.9] &8\\  
HF~LUNG~\cite{hsu2022progressively} &Stethoscope & 4kHz   &lung sound (several breath cycles) &10554 & 15.0[15.0$\sim$15.0]  &8 \\
  \bottomrule     
\end{tabular}
}

\label{tab:data}
\vspace{-10pt}
\end{table}

\begin{figure*}[t]
\begin{adjustbox}{width=1\textwidth,center}
    \centering
    \subfigure[Contrastive (OPERA-CT, OPERA-CE)]{\includegraphics[height=4cm]{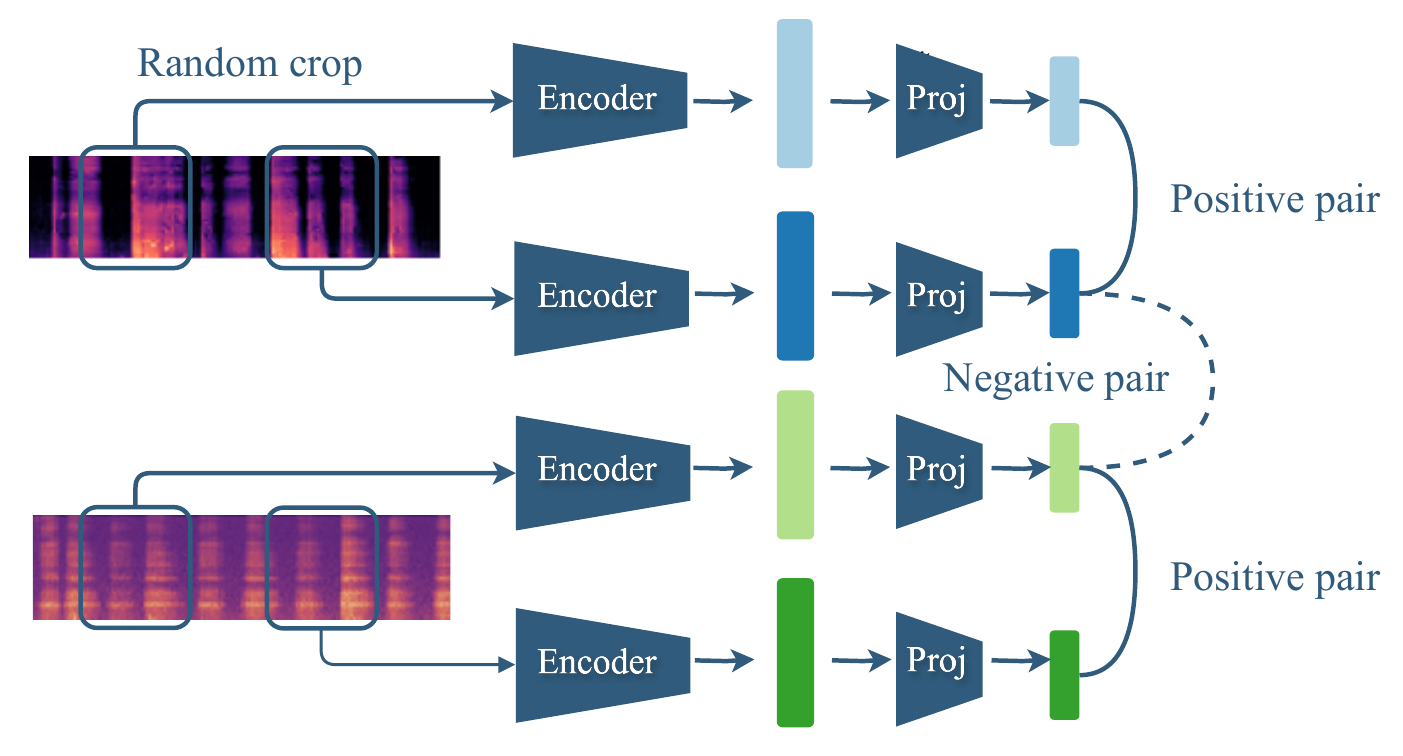}}
    \subfigure[Generative (OPERA-GT)]{ \includegraphics[height=3.8cm]{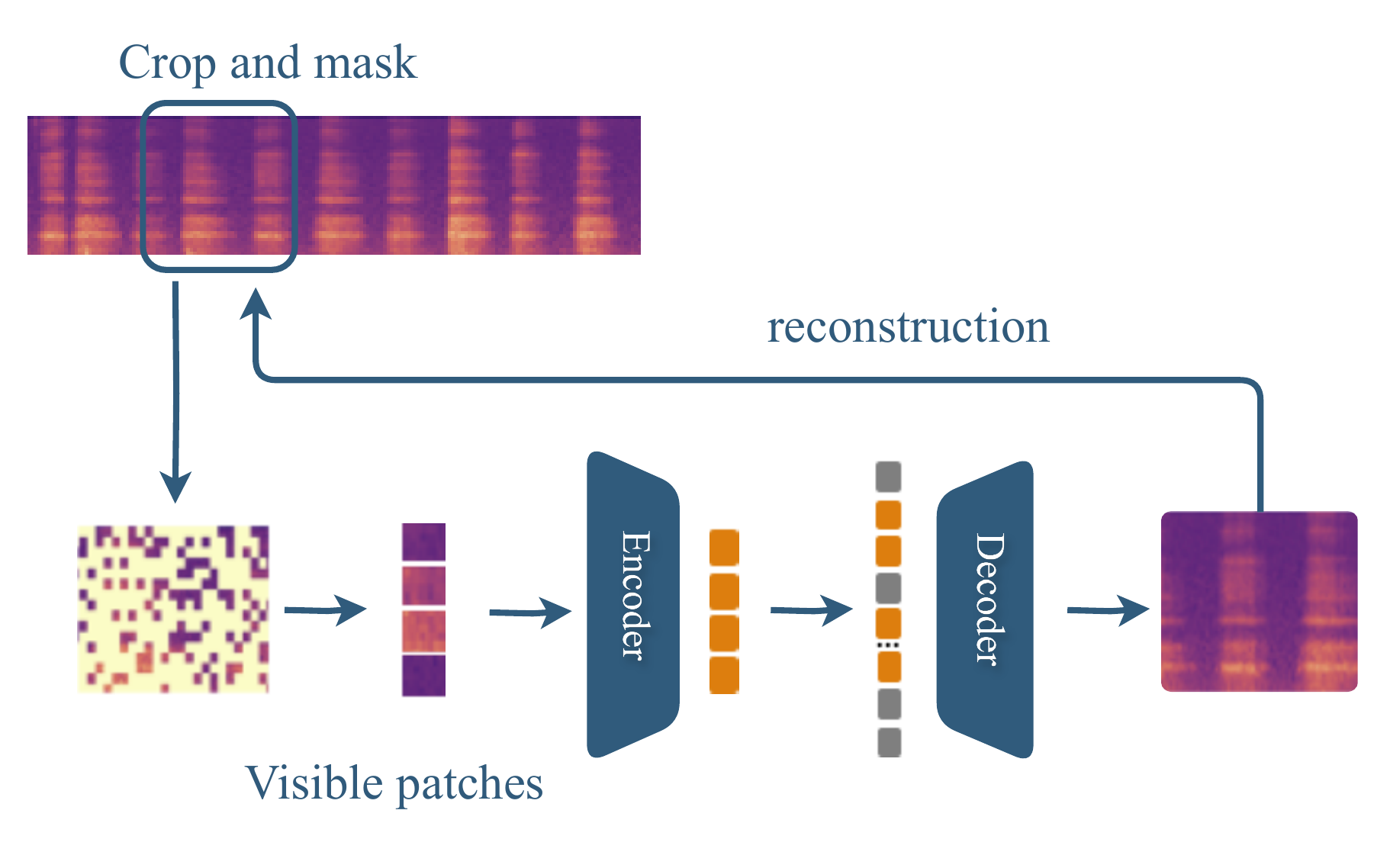}}
     \setlength{\abovecaptionskip}{0.3cm}
\end{adjustbox}
    \caption{Self-supervised learning methods used in our system. }
    \label{fig:pretrain}
\end{figure*}

\subsection{Pretraining Models and Methods}
\vspace{-5pt}

We pre-train our models using a combination of the aforementioned data resources, dividing each dataset into equally-sized batches for consistent processing. We randomly shuffle the batches and reserve 10\% for validation. Due to inherent variations in audio length within individual batches, we employ random cropping of spectrograms, with crop lengths specified in \cref{tab:data}.
Considering the unlabeled nature of the pretraining data, we adopt the most representative SSL methods: contrastive learning-based and generative pretraining-based objectives to pretrain our models. The rationale behind this choice is that if an encoder can distinguish the source of audio segments (contrastive) or reconstruct masked spectrograms (generative), it is expected to have encoded useful and generalizable acoustic features.
The three foundation models we pretrained are: 
\begin{itemize}[topsep=-1pt, itemsep=0pt, leftmargin=15pt]
    \item \textbf{OPERA-CT}: OPERA-CT is a contrastive learning based~\cite{saeed2021contrastive} transformer model. Two segments from the same spectrogram are regard as a positive pair, otherwise negative pairs. As shown in \cref{fig:pretrain}(a), an encoder network (a transformer~\cite{chen2022hts}) extracts features from these segments, and a projector maps them into a low-dimensional representation space, where bilinear similarity is calculated. The optimization objective aims to maximize the similarity between positive pairs and minimize it for negative pairs.
    The encoder has 31M trainable parameters.

     \item \textbf{OPERA-CE}: Similar to OPERA-CT, CE leverages a contrastive pre-training approach. However, it utilizes a more lightweight and efficient CNN encoder (EfficientNet-B0)~\cite{tan2019efficientnet}, which has approximately 4M trainable parameters.

    \item \textbf{OPERA-GT}: OPERA-GT is a generatively pretrained transformer model~\cite{baade2022mae}. As shown in \cref{fig:pretrain}(b),  the encoder (a vision transformer with 21M trainable parameters) is utilized to extract useful features from masked spectrograms, from which the decoder (a lightweight swin-transformer with 12M trainable parameters) can reconstruct the original spectrograms. To train the encoder and the decoder, spectrograms are cropped to equal lengths and then split into small patches.
    We randomly mask 70\% of patches per spectrogram for reconstruction. 

\end{itemize}

Detailed introduction to these three models can be found in  \cref{app:imple}. 
We train them for up to 200 epochs 
and save the best model based on the held-out validation set (i.e., its performance on the pretraining objective). Model checkpoints are also released. More pretraining results and analysis are available in \cref{app:pretrain}.


\section{Benchmarking}\label{sec:benchmark}

\subsection{Benchmark Datasets and Tasks Setup}
\vspace{-5pt}
\begin{table}[t]
\caption{Downstream task characteristics grouped by task category. Datasets in grey are entirely new (not used in pretraining), while others have test sets held out unseen. For T13-T19, FVC denotes forced vital capacity (L), FEV1 is the forced expiratory volume in 1 second, and FEV1/FVC refers to the ratio of the two. 
}
\centering
\begin{adjustbox}{width=\textwidth,center} 
\begin{tabular}{rlllllllll}
 
\toprule
\textbf{Dataset} & \textbf{ID}    & \textbf{Task}             & \textbf{Modality} & \textbf{\#Sam. (\#Sub.)} & \textbf{Data Distribution} \\
\midrule
\textbf{UK~COVID-19}~\cite{coppock2024audio}      & T1 & Covid / Non-covid         & Exhalation             & 2500  (2500)        & 840 / 1660                         \\
                         & T2 & Covid / Non-covid           & Cough             & 2500  (2500)        & 840 / 1660                         \\
\textbf{COVID-19 Sounds}~\cite{xia2021covid}   & T3 & Symptomatic / Healthy       & Breath             & 4138 (3294)        & 2029 / 2109                    \\
                         & T4 & Symptomatic / Healthy       & Cough             & 4138 (3294)       & 2029 / 2109                           \\
\textbf{CoughVID}~\cite{orlandic2021coughvid}         & T5 & Covid / Non-covid           & Cough             & 6175  (n/a)       & 547 / 5628                             \\
                         & T6 & Female / Male               & Cough             & 7263 (n/a)       & 2468 / 4795                            \\
\textbf{ICBHI}~\cite{rocha2019open}           & T7 & COPD / Healthy              & Lung sounds             & 828 (90)        & 793 / 35             \\
{\cellcolor{mygrey}}\textbf{Coswara}~\cite{bhattacharya2023coswara}          & T8 & Smoker / Non-smoker         & Cough             & 948  (n/a)        & 201 / 747                              \\
  {\cellcolor{mygrey}}                       & T9 & Female / Male               & Cough             & 2496  (n/a)       & 759 / 1737                  \\
{\cellcolor{mygrey}}\textbf{KAUH}~\cite{fraiwan2021dataset}           & T10 & Obstructive / Healthy       & Lung sounds             & 234 (79)        & 129 / 105                  \\
{\cellcolor{mygrey}}\textbf{Respiratory@TR}~\cite{altan2017multimedia}                & T11 & COPD severity             & Lung sounds            & 504 (42)        & 72 / 60 / 84 / 84 / 204                   \\
{\cellcolor{mygrey}}\textbf{SSBPR}~\cite{xiao2023snoring}           & T12 & Body position recognition & Snoring             & 7468 (20)       & 1638 / 1454 / 1269 / 1668 / 1439             \\

\midrule



{\cellcolor{mygrey}}\textbf{MMlung}~\cite{mosuilymmlung}           & T13 & FVC                       & Deep breath          & 40 (40)         & 3.402 $\pm$ 1.032 L                    \\
  {\cellcolor{mygrey}}                       & T14 & FEV1                      & Deep breath           & 40   (40)          & 2.657 $\pm$ 0.976 L                  \\
   {\cellcolor{mygrey}}                      & T15 & FEV1/FVC                  & Deep breath             & 40   (40)          & 0.808 $\pm$ 0.190 L                       \\
    {\cellcolor{mygrey}}                     & T16 & FVC                       & O   Vowels         & 40   (40)          & 3.402 $\pm$ 1.032 L                \\
    {\cellcolor{mygrey}}                     & T17 & FEV1                      & O   Vowels              & 40    (40)         & 2.657 $\pm$ 0.976 L                         \\
    {\cellcolor{mygrey}}                     & T18 & FEV1/FVC                  & O   Vowels               & 40   (40)          & 0.808 $\pm$ 0.190 L                  \\
{\cellcolor{mygrey}}\textbf{NoseMic}~\cite{butkow2024evaluation}          & T19 & Respiratory rate          & Breath             & 1297 (16)     & 13.915 $\pm$ 3.386 bpm                   \\

  \bottomrule     
\end{tabular}
\vspace{-20pt}
\end{adjustbox}
\label{tab:task}
\end{table}

\textbf{Tasks}. 
To facilitate the evaluation of our pretrained models, existing acoustic models, and future emerging respiratory acoustic foundation models, we introduce a new benchmark. A total of 10 labeled respiratory audio datasets, encompassing 6 respiratory audio modalities, are curated for this benchmark. Among these 10 datasets, 6 are new and unseen during the pretraining stage.

Using these 10 datasets, we formulate 19 downstream tasks: 12 for health condition inference and 7 for  lung function estimation. The first group covers disease detection such as COVID-19 and COPD (Chronic Obstructive Pulmonary Disease),
participant attribute inference like smoker and gender, disease severity classification, and body position in sleep monitoring. 
Tasks 1-10 are binary classification, while Tasks 11-12 involve 5 classes. The second group includes spirometry test performance and respiratory rate estimation, which are regression tasks aimed at predicting continuous values.
Data and task statistics are summarized in \cref{tab:task}, with detailed descriptions and licenses provided in \cref{app:dataset}.

\textit{All data in this benchmark are publicly available or under controlled access procedures}. 
When available, we follow the official train-test split (Tasks 1-4 and 12-18); otherwise, we implement a random participant-independent split to ensure realistic evaluation (Tasks 5-11 and 19). Due to the limited number of participants in Tasks 13-19, we employ leave-one-subject-out evaluation. For all other tasks, we adopt a fixed random train-validation-test split.

\textbf{Baselines}. In addition to our pretrained models, we also include a commonly used acoustic feature set and three open pretrained acoustic models in this benchmark. They are \textbf{Opensmile} ~\cite{eyben2010opensmile} (\textit{Emobase} acoutic feature set), \textbf{VGGish}~\cite{hershey2017cnn} (supervised pretrained), \textbf{AudioMAE}~\cite{huang2022masked} (self-supervised pretrained) and \textbf{CLAP}~\cite{elizalde2023clap} (language-supervised pretrained). We consider these four methods as baselines to be distinguished from our pretrained models. \reb{We also pretrain these architectures with our OPERA data and results can be found in \cref{app:eval}.}

\textbf{Evaluation protocol}. All tasks are evaluated using the standard linear probe protocol~\cite{chen2020simple, saeed2021contrastive, niizumi2021byol}: training a single fully connected layer on top of the representations extracted from the frozen encoder. Linear evaluation focuses on the quality of learned representations and is applicable to some very small datasets. 
\textbf{AUROC} (area under the receiver operating characteristic) is reported for classification (Task 1-12) and \textbf{MAE} (mean absolute error) is reported for regression (Task 13-19). For a comprehensive overall evaluation, we report \textbf{MRR} (mean reciprocal rank)~\cite{shi2012climf} across tasks. 

For baselines, both the data pre-processing and feature extraction strictly follow their official implementation. For our pretrained models, the same audio preprocessing is used as in pretraining. 
We then segment our audio into short frames to feed into our foundation models to extract features, and  use the averaged representation over these frames as the input for the linear layer~\cite{huang2022masked}. An extended description of the implementing details can be found in \cref{app:imple}.
\textit{Note that 
the baselines and our pretrained models are implemented within the same pipeline, 
making our results easy to reproduce and our benchmark ready to use.}




\begin{table}[t]
\caption{Mean reciprocal ranks on task groups (\textcolor{red}{higher} is better). The best model within each group is highlighted in pink and the second-best is highlighted in blue \reb{(p values reported in \cref{app:eval})}.
}
\centering
\resizebox{0.99\textwidth}{!}
{
\begin{tabular}{lcccccccccccc}
\toprule

Task                       & \# & Opensmile & VGGish & AudioMAE & CLAP &  \textbf{OPERA-CT} & \textbf{OPERA-CE} & \textbf{OPERA-GT}    \\ 
\midrule
     
All                        & 19 & 0.2912 & 0.2289  &  0.2489 & 0.3435 & {\cellcolor{mypink}}0.5632 & 0.4412 & {\cellcolor{myblue}}0.5298 \\
\midrule
Health condition inference & 12 & 0.2190     &  0.1714 &   0.2058  &   0.4319  & {\cellcolor{mypink}}0.6944 &   0.4153                            & {\cellcolor{myblue}}0.4569 \\
Lung function estimation   & 7  & 0.4150    & 0.3276  & 0.3228   & 0.1918 & 0.3381                    & {\cellcolor{myblue}}0.4857  & {\cellcolor{mypink}}0.6548  \\


  \bottomrule     
\end{tabular}
}
\label{tab:rank}
\end{table}

\vspace{-5pt}
\subsection{Experimental Results}
\vspace{-5pt}
We report the MRR of different task groups in \cref{tab:rank}, with the detailed reciprocal ranks of all evaluated methods on each task   provided in \cref{app:eval}. 
The performance metrics for each task are summarized in \cref{tab:res:health} and \cref{tab:res:lung}. Our benchmark demonstrates reliability, as our implementation of baselines achieves comparable performance to those reported in the literature (e.g., existing cough-based COVID-19 detection studies report an AUROC of about $0.65$~\cite{coppock2024audio, xia2021covid}, aligning with our baseline results in Task 2).
Through these extensive experimental results, we now answer the following three main research questions (RQs):

\vspace{5pt}
\textbf{RQ1. Can pretraining a \reb{foundational} model with diverse unlabeled respiratory audio data lead to better performance than baselines designed for general audio?}

From results highlighted in \cref{tab:rank}, it is evident that our pretrained respiratory acoustic foundation models outperform both the acoustic feature set and existing general audio pretrained models.
Among them, OPERA-CT and OPERA-GT achieve the highest MMR scores of 0.5632 and 0.5298, respectively. Looking at \textcolor{red}{\checkmark} and \textcolor{red}{*} in \cref{tab:res:health} and~\ref{tab:res:lung}, the best OPERA model outperforms the acoustic feature set on 17 tasks and the baseline pretrained models on 16 tasks out of the 19 evaluated tasks. This provides a clear positive answer to RQ1.
This advantage likely stems from their exposure to {\em large-scale} and {\em heterogeneous} respiratory audio data, showing the power and promise of respiratory audio foundation models for health applications. 



Now let us dive into the task performance at a finer granularity. 
For classification, an AUROC exceeding 0.7 is typically desirable to demonstrate the utility of the extracted features~\cite{fawcett2006introduction}.
When examining the AUROC in \cref{tab:res:health}, OPERA models achieve an AUROC exceeding 0.7 on 6 of the 12 health condition inference tasks (Task 2, 6-7, 9-10, and 12), whereas the best baseline, CLAP, only surpasses this threshold on 3 tasks (Task  7, 9-10, and 12). This indicates that our models better encode health condition-related information from respiratory audio. 
Regarding lung function estimations (regression tasks), 
the model needs to capture the global dynamics from the entire audio sample and lower MAE indicates better performance.  In \cref{tab:res:lung}, our pretrained models reduce the error in FVC estimation using breathing sounds (Task 13),  FEV1/FVC estimation using breathing sounds (Task 15), FVC estimation using vowel sounds (Task 16), FEV1/FVC estimation using vowel sounds (Task 18), and respiratory rate estimation (Task 19), with performance close to baselines on other tasks. Furthermore, OPERA-GT also achieves a lower standard deviation across subjects, 
suggesting better generalizability and robustness to different subjects, which are of great importance for healthcare applications. 



\begin{table}[t]
\caption{AUROC on health condition inference tasks (\textcolor{red}{higher} is better).  The best model for each task is highlighted. We report mean and standard deviation from five independent runs. \textcolor{red}{\checkmark} and \textcolor{red}{*} indicates superiority over the opensmile feature set and the other pretrained baselines respectively.
}
\centering
\begin{adjustbox}{width=0.99\textwidth,center} 
\begin{tabular}{ll|c|ccc|ccc|c}
\toprule
ID & Task Abbr.        & Opensmile     & VGGish        & AudioMAE      & CLAP                                                                    & \textbf{OPERA-CT}  &                                                \textbf{OPERA-CE}                                 & \textbf{OPERA-GT}                  &                                            \\ 
\midrule

T1      & Covid (Exhale)    & 0.550 $\pm$ 0.015 & 0.580 $\pm$ 0.001 & 0.549 $\pm$ 0.001 & 0.565 $\pm$ 0.001                                                           & 0.586 $\pm$ 0.008                                                          & 0.551 $\pm$ 0.010                   & \cellcolor{mypink}{0.605 $\pm$ 0.001} &\textcolor{red}{\checkmark*} \\
T2      & Covid (Cough)     & 0.649 $\pm$ 0.006 & 0.557 $\pm$ 0.005 & 0.616 $\pm$ 0.001 & 0.648 $\pm$ 0.003                                                           & \cellcolor{mypink}{0.701 $\pm$ 0.002}& 0.629 $\pm$ 0.006                                                           & 0.677 $\pm$ 0.001                                &\textcolor{red}{\checkmark*}                            \\
T3      & Symptom (Breath) & 0.571 $\pm$ 0.006 & 0.571 $\pm$ 0.003 & 0.583 $\pm$ 0.003 & 0.611 $\pm$ 0.006                                                           & 0.603 $\pm$ 0.005                                                           & 0.610 $\pm$ 0.004                    & \cellcolor{mypink}{0.613 $\pm$ 0.002}  &\textcolor{red}{\checkmark*}\\
T4      & Symptom (Cough)  & 0.633 $\pm$ 0.012 & 0.605 $\pm$ 0.004 & 0.659 $\pm$ 0.001 & 0.669 $\pm$ 0.002                                                           & \cellcolor{mypink}{0.680 $\pm$ 0.006}& 0.665 $\pm$ 0.001                                                           & 0.673 $\pm$ 0.001                                  &\textcolor{red}{\checkmark*}                          \\
T5      & Covid (Cough)   & 0.537 $\pm$ 0.011	 & 0.538 $\pm$ 0.028  & 	0.554 $\pm$ 0.004 & 	\cellcolor{mypink}{}0.599 $\pm$ 0.007	 & 0.578 $\pm$ 0.001	 & 0.566 $\pm$ 0.008	 & 0.552 $\pm$ 0.003 & \textcolor{red}{\checkmark~~}  \\
T6      & Gender (Cough)   & 0.677 $\pm$ 0.005	 & 0.600 $\pm$ 0.001  & 	0.628 $\pm$ 0.001	 & 0.665 $\pm$ 0.001 & 	\cellcolor{mypink}{}0.795 $\pm$ 0.001	 & 0.721 $\pm$ 0.001	 & 0.735 $\pm$ 0.000 & \textcolor{red}{\checkmark*}   \\

T7      & COPD (Lung)        & 0.579 $\pm$ 0.043 & 0.605 $\pm$ 0.077 & 0.886 $\pm$ 0.017 & \cellcolor{mypink}{0.933 $\pm$ 0.005}& 0.855 $\pm$ 0.012                                                           & 0.872 $\pm$ 0.011                                                           & 0.741 $\pm$ 0.011                                     &\textcolor{red}{\checkmark}~~                       \\
T8      & Smoker (Cough)   & 0.534 $\pm$ 0.060 & 0.507 $\pm$ 0.027 & 0.549 $\pm$ 0.022 & 0.680 $\pm$ 0.009                                                           & \cellcolor{mypink}{0.685 $\pm$ 0.012}& 0.674 $\pm$ 0.013                                                           & 0.650 $\pm$ 0.005                                       &\textcolor{red}{\checkmark*}                     \\
T9      & Gender (Cough)     & 0.753 $\pm$ 0.008 & 0.606 $\pm$ 0.003 & 0.724 $\pm$ 0.001 & 0.742 $\pm$ 0.001                                                           & \cellcolor{mypink}{0.874 $\pm$ 0.000}& 0.801 $\pm$ 0.002                                                           & 0.825 $\pm$ 0.001                                       &\textcolor{red}{\checkmark*}                     \\
T10     & Obstructive (Lung)            & 0.636 $\pm$ 0.082 &	0.605 $\pm$ 0.036 &		0.616 $\pm$ 0.041 &		0.697 $\pm$ 0.004	&	0.722 $\pm$ 0.016	&	\cellcolor{mypink}0.741 $\pm$ 0.014	&	0.703 $\pm$ 0.016 &\textcolor{red}{\checkmark*}  \\
T11     & COPD severity (Lung)     & 0.494 $\pm$ 0.054 & 0.590 $\pm$ 0.034 & 0.510 $\pm$ 0.021 & 0.636 $\pm$ 0.045                                                           & 0.625 $\pm$ 0.038                                                           & \cellcolor{mypink}{0.683 $\pm$ 0.007}& 0.606 $\pm$ 0.015                                  &\textcolor{red}{\checkmark*}                          \\
T12     & Position (Snoring)       & 0.772 $\pm$ 0.005 & 0.657 $\pm$ 0.002 & 0.649 $\pm$ 0.001 & 0.702 $\pm$ 0.001                                                           & \cellcolor{mypink}{0.781 $\pm$ 0.000}& 0.769 $\pm$ 0.000                                                           & 0.742 $\pm$ 0.001                                  &\textcolor{red}{\checkmark*}                          \\

  \bottomrule     
\end{tabular}
\end{adjustbox}
\label{tab:res:health}
\end{table}


\begin{table}[t]
\caption{MAE on lung function estimation tasks (\textcolor{red}{lower} is better). Best model per task is highlighted. We report mean and standard deviation across subjects. }
\centering
\begin{adjustbox}{width=0.99\textwidth,center} 
\begin{tabular}{ll|c|ccc|cccccc|c}
\toprule
ID & Task Abbr.  & Opensmile     & VGGish        & AudioMAE      & CLAP          & \textbf{OPERA-CT} & \textbf{OPERA-CE} & \textbf{OPERA-GT} &\\
  \midrule
T13 & FVC (Breath)          &0.985 $\pm$ 0.743	&0.904 $\pm$ 0.568	&0.900 $\pm$ 0.551	&0.896 $\pm$ 0.542	&0.924 $\pm$ 0.583	&\cellcolor{mypink}0.848 $\pm$ 0.607	&0.892 $\pm$ 0.618 &\textcolor{red}{\checkmark*}\\
T14 & FEV1 (Breath)         & \cellcolor{mypink}0.756 $\pm$ 0.721	&0.839 $\pm$ 0.563	&0.821 $\pm$ 0.590	&0.840 $\pm$ 0.547	&0.837 $\pm$ 0.563	&0.834 $\pm$ 0.581	&0.825 $\pm$ 0.560 &  \\
T15 & FEV1/FVC (Breath)     & 0.141 $\pm$ 0.185	&0.131 $\pm$ 0.146	&0.129 $\pm$ 0.146	&0.134 $\pm$ 0.146	&\cellcolor{mypink}0.128 $\pm$ 0.140	&0.132 $\pm$ 0.141	&\cellcolor{mypink}0.128 $\pm$ 0.141 &\textcolor{red}{\checkmark*}\\
T16 & FVC (Vowel)           & 0.850 $\pm$ 0.592	&0.895 $\pm$ 0.559	&0.833 $\pm$ 0.588	&0.883 $\pm$ 0.560	&0.885 $\pm$ 0.553	&\cellcolor{mypink}0.761 $\pm$ 0.544	&0.878 $\pm$ 0.550  &\textcolor{red}{\checkmark*}  \\
T17 & FEV1 (Vowel)           &\cellcolor{mypink}0.730 $\pm$ 0.497	&0.842 $\pm$ 0.559	&0.876 $\pm$ 0.561	&0.859 $\pm$ 0.541	&0.780 $\pm$ 0.542	&0.830 $\pm$ 0.561	&0.774 $\pm$ 0.554    &\textcolor{red}{~~~*}  \\
T18 & FEV1/FVC (Vowel)        &0.138 $\pm$ 0.166	&\cellcolor{mypink}0.130 $\pm$ 0.145	&0.131 $\pm$ 0.141	&0.137 $\pm$ 0.147	&0.132 $\pm$ 0.140	&0.136 $\pm$ 0.150	&\cellcolor{mypink}0.130 $\pm$ 0.138 &\textcolor{red}{\checkmark*}  \\
T19 & Breathing Rate        &2.714 $\pm$ 0.902	&2.605 $\pm$ 0.759	&2.641 $\pm$ 0.813	&2.650 $\pm$ 0.947	&2.636 $\pm$ 0.858	&2.525 $\pm$ 0.782	&\cellcolor{mypink}2.416 $\pm$ 0.885 &\textcolor{red}{\checkmark*} \\

  \bottomrule     
\end{tabular}
\end{adjustbox}
\label{tab:res:lung}
\vspace{-5pt}
\end{table}

\vspace{5pt}
\textbf{RQ2. Are the pretrained respiratory acoustic models generalizable to new data?}

It is crucial that foundation models can generalize to new and unseen data once developed. In our benchmark, we have 12 tasks formulated from unseen datasets (Task 8-19) and unseen respiratory audio modalities (Task 12, 16-18) not used for pretraining. Notably, our respiratory acoustic foundation models demonstrate good generalization capabilities, achieving the best performance on 5 out of 5 classification tasks and 5 out of 7 regression tasks. They are able to outperform the acoustic feature set and general audio pretrained models which are supposed to exhibit generalizability. Specifically, in \cref{tab:res:health}, Task 8-12 all have an AUROC higher than 0.68. Comparing Task 6 and Task 9 with the same prediction target, the performance on unseen data (Task 9) is comparable. Therefore, our foundation models are generalizable, likely due to the minimal assumptions made during SSL pretraining. \reb{We have additional experiments on cross-domain zero-shot performance in \cref{app:eval}.}


\vspace{5pt}
\textbf{RQ3. How to design SSL methods and model architectures of the pretrained respiratory acoustic encoders with different applications in mind?}

Within the OPERA system, we train foundation models using two different SSL strategies: contrastive and generative. 
From \cref{tab:rank}, \ref{tab:res:health}, and \ref{tab:res:lung}, it can be observed that the models pretrained with a contrastive objective (OPERA-CT, OPERA-CE) generally achieve superior performance on classification tasks (i.e., health condition inference),  while the generative pretrained models (OPERA-GT and baseline AudioMAE) perform better on regression tasks (i.e., lung function estimation). This finding aligns with the inherent nature of the methods, as contrastive learning's discriminative training goal naturally aligns with the classification objective, and it discards the decoder in the architecture compared to generative models.
It is also consistent with prior observations on various vision benchmarks~\cite{liu2021self}.

\begin{figure}[t]
    \centering
    \subfigure[Saliency maps for T2 COVID detection with a spectrogram of three coughs. OPERA-CT highlights strong gradients for  low and high-frequency bins in the middle while GT spreads attention across various time frames and frequency bins, less focused on the coughs.]{
    \includegraphics[width=0.48\textwidth]{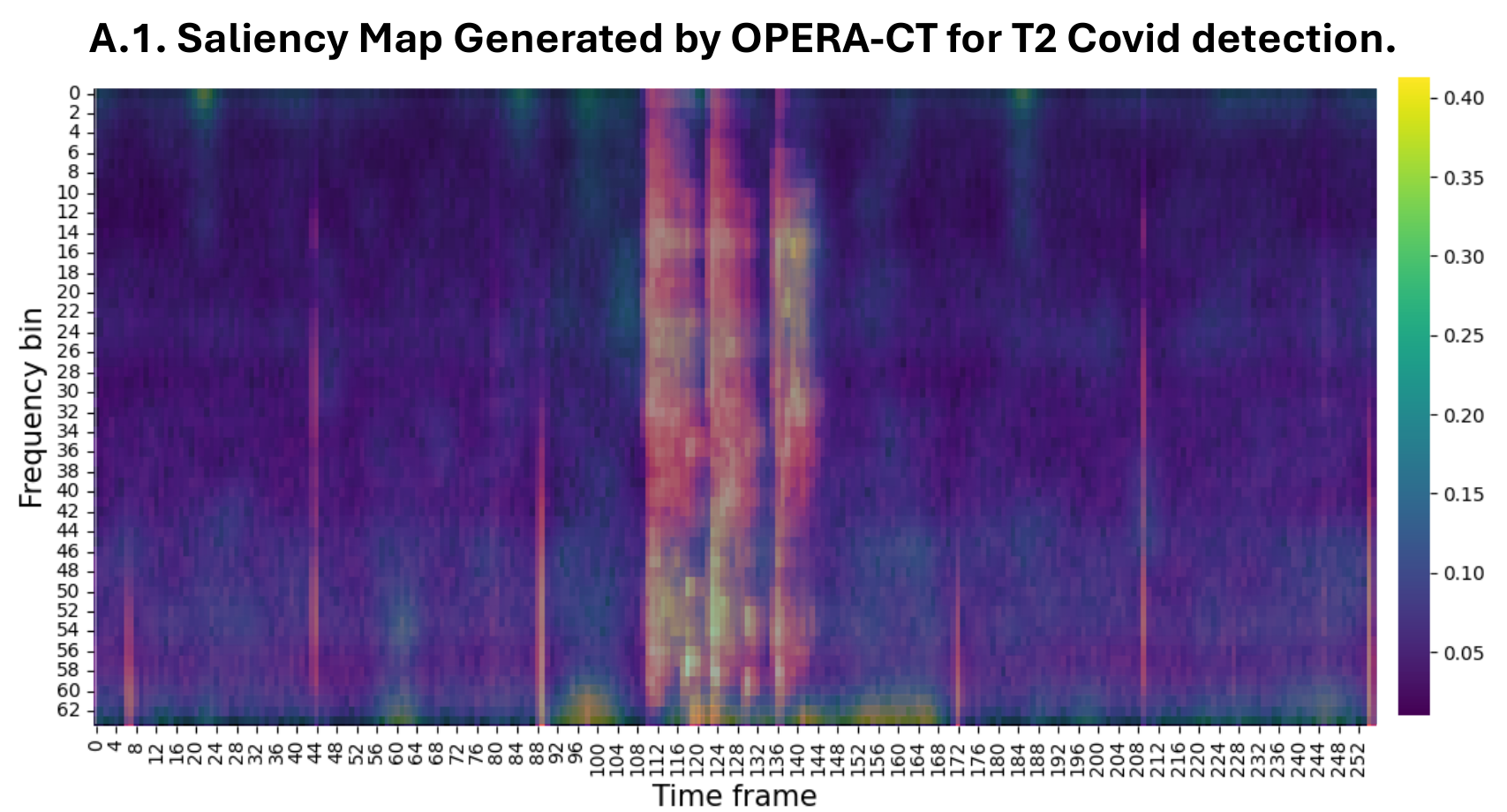}
    \includegraphics[width=0.48\textwidth]{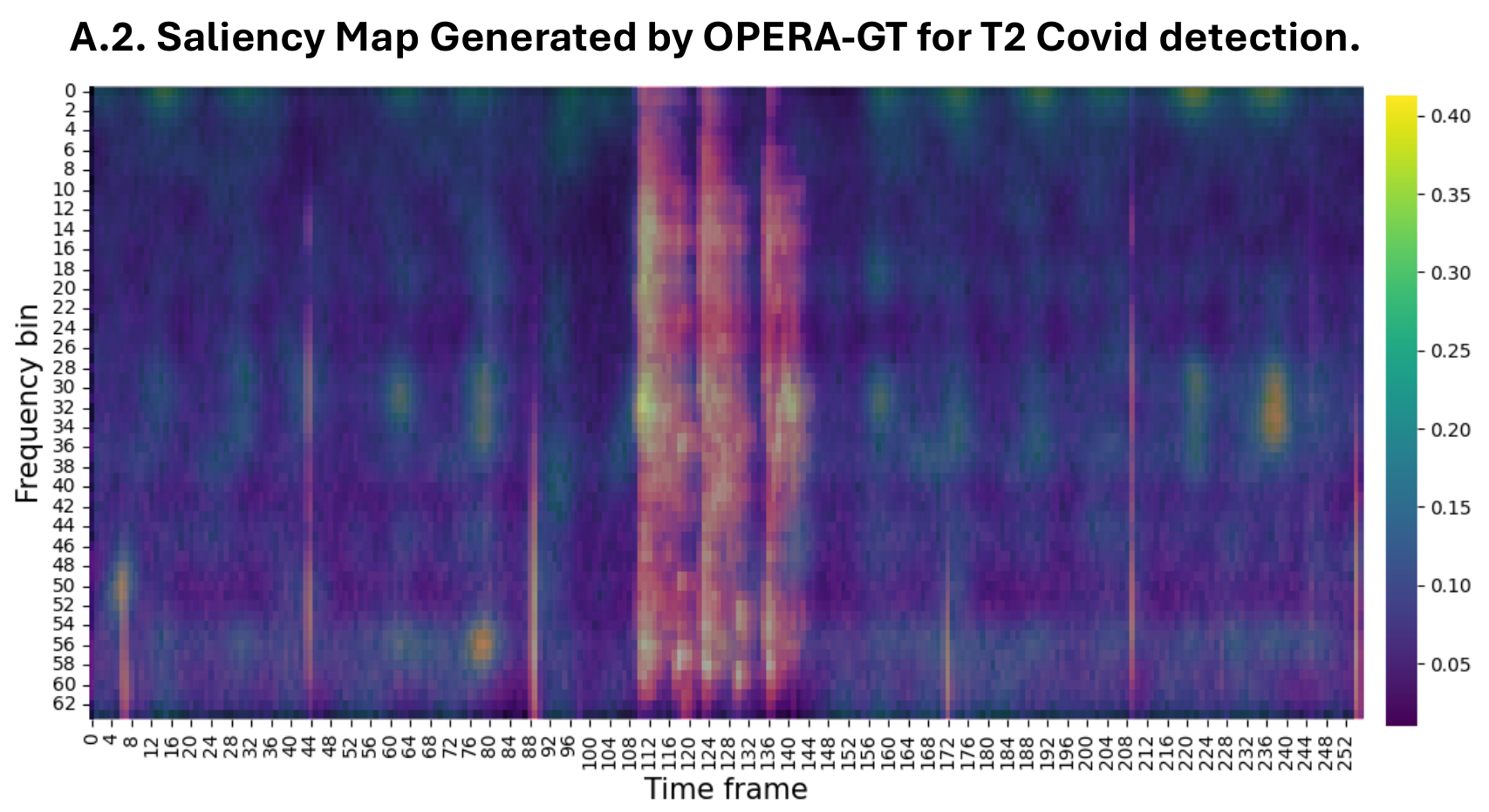}
    }
    \subfigure[Saliency maps for T13 FVC estimation with a spectrogram of one respiratory cycle. OPERA-CT highlights high-frequency bins in the upper-right corner, while GT focuses on decayed energy in high-frequency bins, which is more useful for this task.]{
    \includegraphics[width=0.48\textwidth]{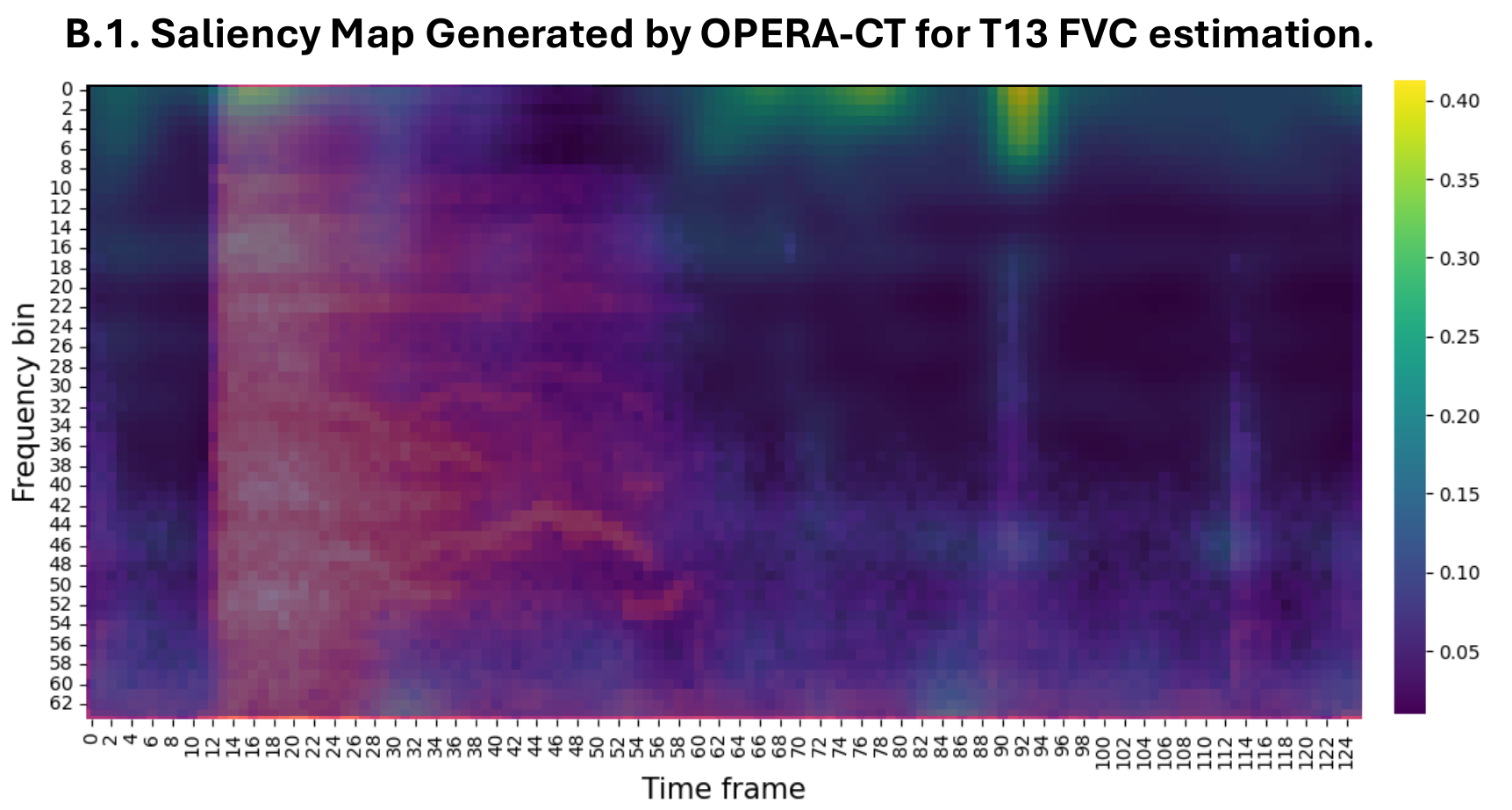}
    \includegraphics[width=0.48\textwidth]{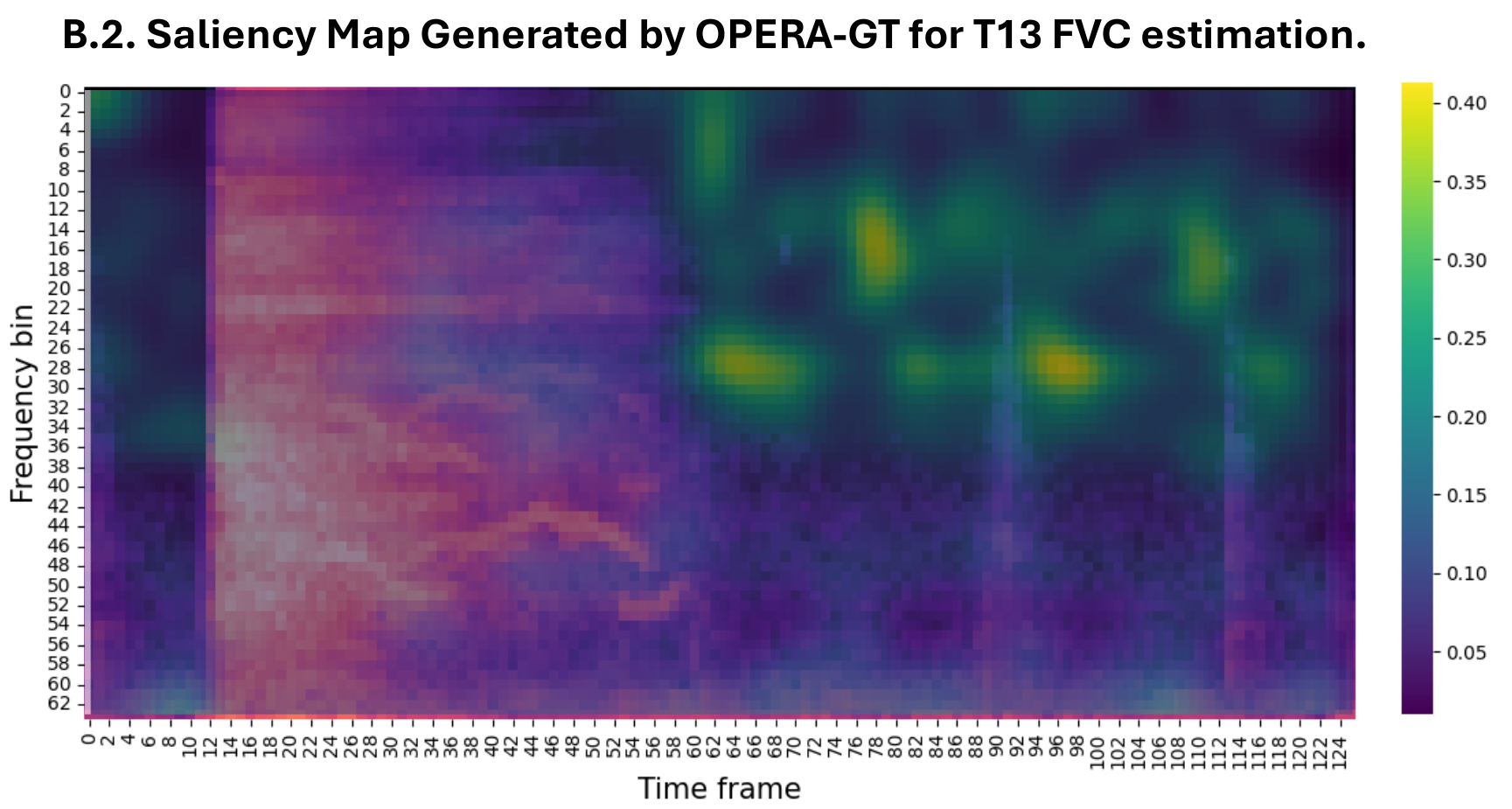}
    }
    \subfigure[Saliency maps for T13 FVC estimation with a spectrogram of four respiratory cycles. Both models highlight the time frames with silence which are useful for this task.]{
    \includegraphics[width=0.48\textwidth]{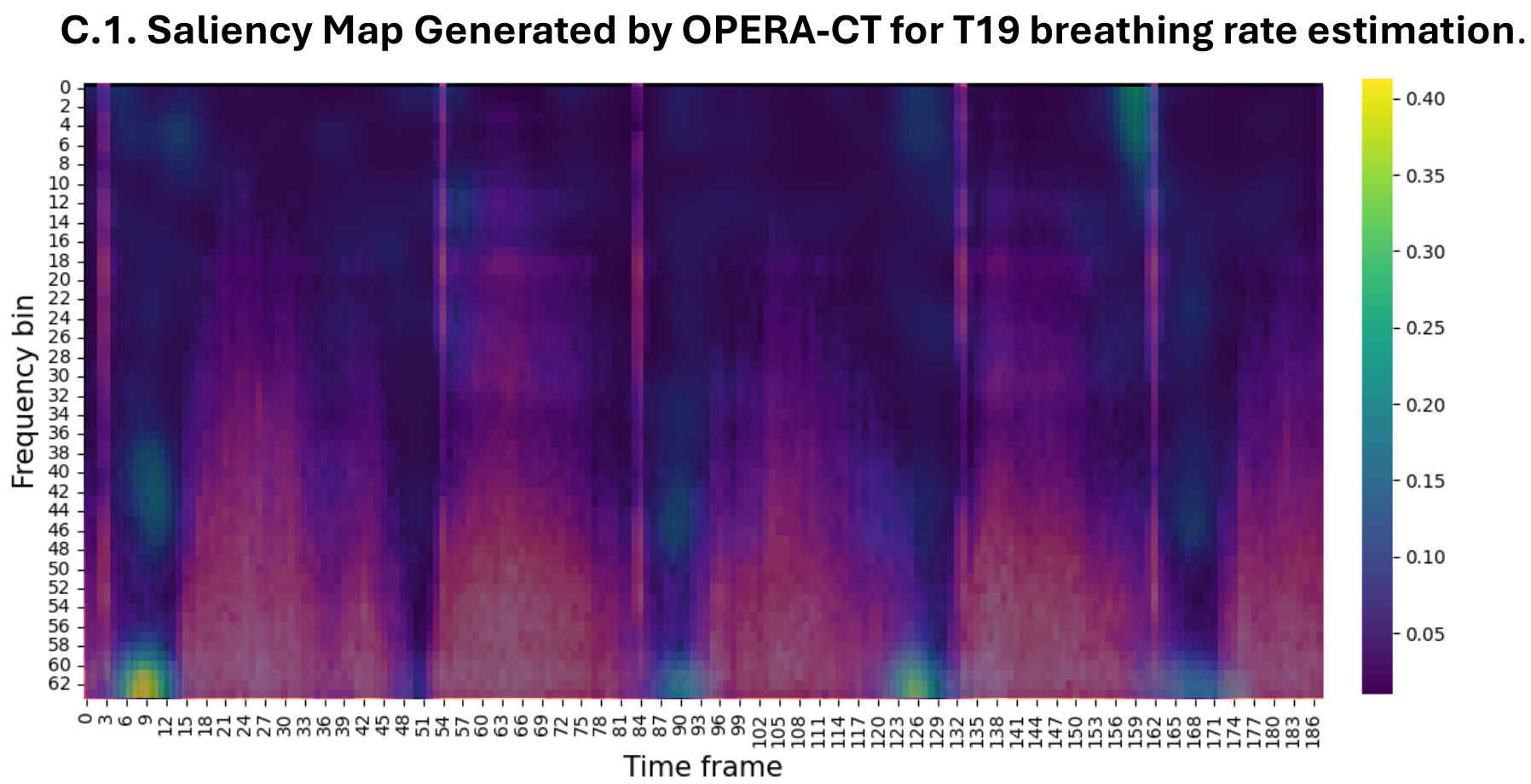}
    \includegraphics[width=0.48\textwidth]{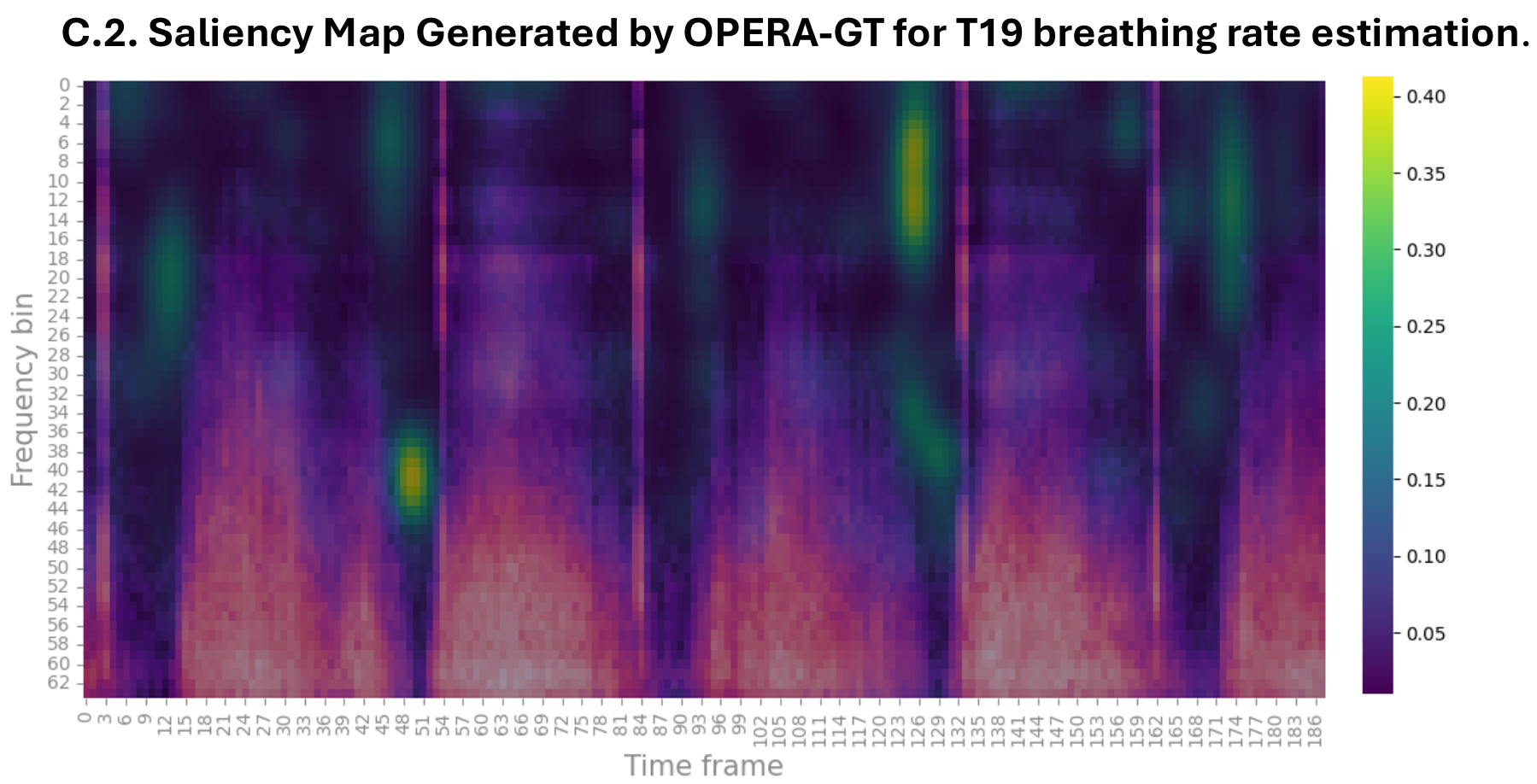}
    }
    \caption{Saliency maps generated by OPERA-CT and OPERA-GT on three example tasks (T2, T13, and T19). The yellow color indicates the largest gradient on the spectrogram.
    }
    \label{fig:saliency}
\end{figure}

\reb{To gain deeper insight, we further use saliency maps to explicitly inspect what our models focus on in the spectrograms of unseen audio data. \cref{fig:saliency} presents examples for three tasks. We observe that the OPERA-CT model tends to focus on a few local areas of the spectrogram, showing distinct saliency peaks, whereas the GT model analyzes the global distribution with a more even saliency map. Comparing Figures~\ref{fig:saliency} B.1 and B.2, GT demonstrates greater ability in detecting the continuous decline in energy after breathing. This explains why CT underperforms compared to the GT variant in lung function estimation, where global patterns are more critical. Based on this observation, we explored a hybrid model that combines both pretraining objectives. However, results show no consistent improvements over the individual objectives. Detailed results are provided in \cref{app:eval}. This suggests that such a simple combination is not sufficient, leaving space for further exploration.}

We also compare CNN and transformer encoder architectures using the same SSL strategy.
Overall, our results suggest a strong representation ability of the transformer architecture for audio.
Specifically, OPERA-CT performs the best in 7 out of the 12 health condition inference tasks (\cref{fig:radar:health}), with a mean reciprocal rank as high as 0.6944 (\cref{tab:rank}). For lung function estimation tasks, OPERA-GT performs the best in 3 out of the 7 tasks  (\cref{fig:radar:lung}), with the highest mean reciprocal rank of 0.6548 (\cref{tab:rank}) and achieves the second on health condition inference tasks. 
As a lightweight CNN model, OPERA-CE also demonstrates satisfactory results, with a mean reciprocal of 0.4690, and performs third and second best in the two groups of tasks respectively (\cref{tab:rank}). This shows the promise of training a lightweight foundation model for efficient computing and on-device learning for resource-constrained scenarios.

\section{Conclusion and Future Research Directions} 
\label{sec:discussion}

In this paper,  we present \textit{OPERA}, the first open-source respiratory acoustic foundation model pretraining and benchmarking system. 
OPERA offers a unique curated dataset pool, a ready-to-use evaluation portal as well as a thorough analysis of performance across architectures and tasks.  We discuss the limitations of our work and how it can serve as a foundation for future explorations.

\paragraph{Limitations.}
\reb{While our benchmark is comprehensive, covering 19 tasks, we note that some labels, such as the COVID-19 test results in Task 5, were obtained and reported by participants themselves. As a result, some labels could be less precise than clinically validated data like Task 2. This issue does not affect the pretrained foundation models, as they are trained without these labels. 
Additionally, OPERA is \textit{not} intended for clinical use and should not be considered safe for such applications. Care should be taken to prevent potential misuse when utilizing the models.}

\reb{In addition to the study we have done, OPERA can support a number of future explorations: }

\textbf{(1) Studying data-efficient fine-tuning}. 
\cref{sec:benchmark} uses linear evaluation with frozen encoders following standard protocols and accommodating limited downstream data (see \cref{tab:task}).  
We select some tasks with relatively abundant labeled data to examine fine-tuning performance (details  in \cref{app:eval}). Results for Task 4 are presented in \cref{tab:finetune}.  
Using the same number of labeled data as in linear probing (1749 samples),  all models show improved performance and the three OPERA models achieve an AUROC above $0.7$. With more labeled data for fine-tuning (6648 samples), the best OPERA-GT model achieves an AUROC of  $0.739$. Similarly, OPERA-CT's performance on Task 12 (7468 samples) could be enhanced to $0.994$ compared to $0.781$ in linear evaluation.



\begin{table}[t]
\centering

\caption{AUROC (\textcolor{red}{higher} is better) for linear probing and finetuning on T4. Best model highlighted.}
\label{tab:finetune}
\begin{adjustbox}{width=0.9\textwidth,center} %
\begin{tabular}{lllllll} 
\toprule
   Method   & \# Train & AudioMAE  & CLAP  & OPERA-CT  & OPERA-CE  & OPERA-GT  \\

 \midrule
\textbf{Linear}             & 1749                      & 0.659 $\pm$ 0.001           & 0.669 $\pm$ 0.002           & \cellcolor{mypink}{0.680 $\pm$ 0.006}  & 0.665 $\pm$ 0.001             & 0.673 $\pm$ 0.001            \\
\textbf{Fine-tune}           & 1749                      & 0.672 $\pm$ 0.039           & 0.691 $\pm$ 0.008           & {0.710 $\pm$ 0.003}  & 0.703 $\pm$ 0.003             &        \cellcolor{mypink}{}0.715 $\pm$ 0.006                 \\
\textbf{Fine-tune} & 6648                      & 0.723 $\pm$ 0.010           & 0.723 $\pm$ 0.009           & \cellcolor{mypink}{0.739 $\pm$ 0.008}  &    0.733 $\pm$ 0.002                       &      0.735 $\pm$ 0.005                   \\
\bottomrule
\end{tabular}\label{tab:finetune}
\end{adjustbox}
\vspace{-5pt}
\end{table}

However, most other tasks have a much smaller training set, and thus data efficient large model fine-tuning approaches are desirable. Methods have been proposed in the machine learning literature such as adapter tuning~\cite{hu2023llm}, prefix tuning~\cite{vos2022towards}, prompt tuning~\cite{fathullah2024prompting}, and low-rank adaptation~\cite{hu2021lora}. Yet, they are not designed for audio (spectrograms) or acoustic foundation models.  Considering the properties of downstream health-related tasks which often exhibit limited and imbalanced data, novel audio-specific data efficient fine tuning methods need to be explored. 





\textbf{(2) Investigating scaling law in respiratory acoustic foundation models.}
Recent research on foundation models has uncovered their emergent abilities, largely arising from scaling up pretraining data and model size~\cite{schaeffer2024emergent}. It is also interesting to study the scaling laws in respiratory acoustic foundation models. \reb{While the OPERA dataset is already extensive, further expansion would be valuable for this purpose.} Our benchmark can help to quantify how increasing a model's scale and its training data can significantly enhance performance on downstream tasks.
Based on the currently 404 hours of respiratory audio, our OPERA-CT (31M parameters) and OPERA-GT (21M) models  surpass the lightweight OPERA-CE model (4M). 
With the rapid accumulation of respiratory audio datasets~\cite{xia2022exploring, dang2023human}, more evaluation of scaling laws should
be conducted in future.

\textbf{(3) Exploring novel pretraining strategies for unlabeled health audio.}
We have pretrained three models (OPERA-CT, OPERA-GT, OPERA-CE) and compared their performance. More configurations in terms of model size, architecture, and pretraining methods could be compared in the future. 
Among the two representative SSL approaches we adapted for pretraining, there exist limitations: For contrastive learning, defining positive and negative pairs is challenging due to downstream task diversity, and our definitions might not be optimal. In generative pretraining, using alternative objectives to reconstruction might improve performance on discriminative tasks. 
Combining these methods could be beneficial but presents challenges in balancing objectives, and previous studies suggest simple combinations do not improve performance~\cite{baade2022mae}.
Audio data also pose unique challenges like heterogeneous sound types, varying sampling rates and durations, and complex temporal-frequency correlations, requiring tailored solutions to better pretrain and apply the foundation models. 
OPERA provides a framework for exploring these technical challenges. 


 By introducing this open-source system, we hope to lay the groundwork for responsible, reliable, and sustainable development of foundation models in respiratory healthcare, paving the way for a healthier future for generations to come.  

\begin{ack}
This work was supported by ERC Project 833296 (EAR), EPSRC Project RELOAD,  Nokia Bell Labs through a donation, Cambridge Trust, and the Institute for Life
Sciences (IfLS) HEIF Research Stimulus Fund.
\end{ack}



{
\small



%

\bibliographystyle{ACM-Reference-Format}
\pagenumbering{gobble}

\bibliography{bibliography}
}

\newpage
\section*{Checklist}


\begin{enumerate}

\item For all authors...
\begin{enumerate}
  \item Do the main claims made in the abstract and introduction accurately reflect the paper's contributions and scope?
    \answerYes{}
  \item Did you describe the limitations of your work?
    \answerYes{} See \cref{sec:discussion}.
  \item Did you discuss any potential negative societal impacts of your work?
    \answerNo{} We did not identify any.
  \item Have you read the ethics review guidelines and ensured that your paper conforms to them?
     \answerYes{}
\end{enumerate}

\item If you are including theoretical results...
\begin{enumerate}
  \item Did you state the full set of assumptions of all theoretical results?
    \answerNA{}
	\item Did you include complete proofs of all theoretical results?
    \answerNA{}
\end{enumerate}

\item If you ran experiments (e.g. for benchmarks)...
\begin{enumerate}
  \item Did you include the code, data, and instructions needed to reproduce the main experimental results (either in the supplemental material or as a URL)?
    \answerYes{}
  \item Did you specify all the training details (e.g., data splits, hyperparameters, how they were chosen)?
    \answerYes{}
	\item Did you report error bars (e.g., with respect to the random seed after running experiments multiple times)?
    \answerYes{}
	\item Did you include the total amount of compute and the type of resources used (e.g., type of GPUs, internal cluster, or cloud provider)?
    \answerYes{} See \cref{app:imple}.
\end{enumerate}

\item If you are using existing assets (e.g., code, data, models) or curating/releasing new assets...
\begin{enumerate}
  \item If your work uses existing assets, did you cite the creators?
    \answerYes{}
  \item Did you mention the license of the assets?
    \answerYes{} See \cref{app:dataset}.
  \item Did you include any new assets either in the supplemental material or as a URL?
    \answerYes{} 
  \item Did you discuss whether and how consent was obtained from people whose data you're using/curating?
    \answerYes{} See \cref{app:dataset}.
  \item Did you discuss whether the data you are using/curating contains personally identifiable information or offensive content?
    \answerYes{} See \cref{app:dataset}.
\end{enumerate}

\item If you used crowdsourcing or conducted research with human subjects...
\begin{enumerate}
  \item Did you include the full text of instructions given to participants and screenshots, if applicable?
\answerNA{}
  \item Did you describe any potential participant risks, with links to Institutional Review Board (IRB) approvals, if applicable?
  \answerNA{}
  \item Did you include the estimated hourly wage paid to participants and the total amount spent on participant compensation?
   \answerNA{}
\end{enumerate}

\end{enumerate}


\appendix

\newpage
\pagenumbering{roman}
\section{Appendix for OPERA}
\MiniToC

\subsection{Datasets Overview}
\label{app:dataset}





We have used 11 datasets in our benchmark. Their statistics are summarized in \cref{tab:data} and  \cref{tab:task} in the main paper. Here, we supplement their access methods and licenses in \cref{tab:data_url} with a more detailed description below. It can be noted that all datasets contain an audio set and a metadata part. Audio data used are anonymous and the metadata do not contain personally identifiable information or offensive content. 

\textbf{COVID-19 Sounds~\cite{xia2021covid} }. The COVID-19 Sounds dataset consists of 53,449 audio samples (over 552 hours in total) crowd-sourced from 36,116 participants through the COVID-19 Sounds app. This dataset is comprehensive in terms of demographics and spectrum of health conditions. It also provides participants' self-reported COVID-19 testing status with 2,106 samples tested positive. It consists of three modalities including breathing, cough, and voice recordings. Only breathing and cough modalities are used in this paper.

This dataset is crowdsourced through the COVID-19 Sounds project, approved by the Ethics Committee of the Department of Computer Science and Technology at the University of Cambridge. Informed consent was obtained from all the participants. The dataset is accessible under controlled access through a Data Transfer Agreement and has been widely shared and used~\cite{xue2021exploring, rizos2023positive}.

\textbf{UK COVID-19~\cite{coppock2024audio}}. The UK COVID-19 Vocal Audio Dataset is designed for the training and evaluation of machine learning models that classify SARS-CoV-2 infection status or associated respiratory symptoms using vocal audio. The UK Health Security Agency recruited voluntary participants through the national Test and Trace programme and the REACT-1 survey in England from March 2021 to March 2022, during dominant transmission of the Alpha and Delta SARS-CoV-2 variants and some Omicron variant sublineages. Audio recordings of volitional coughs, exhalations, and speech (speech not included in open access version, nor used in this paper) were collected in the `Speak up to help beat coronavirus' digital survey alongside demographic, self-reported symptom and respiratory condition data, and linked to SARS-CoV-2 test results. 


The study has been approved by The National Statistician’s Data Ethics Advisory Committee (reference NSDEC(21)01) and the Cambridge South NHS Research Ethics Committee (reference 21/EE/0036) and Nottingham NHS Research Ethics Committee (reference 21/EM/0067).
Participants reviewed the participant information and confirmed their informed consent to take part.

\textbf{COUGHVID~\cite{orlandic2021coughvid}}. The COUGHVID dataset provides over 25,000 crowdsourced cough recordings representing a wide range of participant ages, genders, geographic locations, and COVID-19 statuses.

All of the data collection and annotation was done in compliance with relevant ethical regulations. Informed consent was obtained by all participants who uploaded their cough sounds and metadata.

\textbf{ICBHI~\cite{rocha2019open}}.
The ICBHI Respiratory Sound Database contains audio samples, collected independently by two research teams in two different countries, over several years. Ethical approval was obtained from the ethics committees of the appropriate institutions.

Most of the database consists of audio samples recorded by the School of Health Sciences, University of Aveiro (ESSUA) research team at the Respiratory Research and Rehabilitation Laboratory (Lab3R), ESSUA and at Hospital Infante D. Pedro, Aveiro, Portugal. The second research team, from the Aristotle University of Thessaloniki (AUTH) and the University of Coimbra (UC), acquired respiratory sounds at the Papanikolaou General Hospital, Thessaloniki and at the General Hospital of Imathia (Health Unit of Naousa), Greece.
The database consists of a total of 5.5 hours of recordings in 920 annotated audio samples from 126 subjects.



\begin{table}
\centering
\caption{Dataset availability. *ICBHI and HF Lung datasets coming from multiple sources, please refer to the text description below. COVID-19 Sounds, SSBPR, MMLung and NoseMic are available upon request. The custom license is detailed in the DTA (data transfer agreement). }
\label{tab:license}
\begin{adjustbox}{width=\textwidth,center} 
\begin{tabular}{llll} 
\toprule
Dataset                                                                                           & Source & Access & license  \\

\midrule
COVID-19 Sounds\cite{xia2021covid}  & UoC                                      &  \url{https://covid-19-sounds.org/blog/neurips_dataset}    & Custom license     \\
UK COVID-19~\cite{coppock2024audio}  & IC  &   \url{https://zenodo.org/records/10043978}   &     OGL 3.0      \\
CoughVID\cite{orlandic2021coughvid}  &  EPFL & \url{https://zenodo.org/records/4048312}  &  CC BY 4.0 \\                                  
ICBHI\cite{rocha2019open}                                                  &   *   &     \url{https://bhichallenge.med.auth.gr}    &  CC0 \\

HF Lung~\cite{hsu2022progressively}         &  *    &   \url{https://gitlab.com/techsupportHF/HF_Lung_V1} &  CC BY 4.0   \\

& & \url{https://gitlab.com/techsupportHF/HF_Lung_V1_IP}  &  CC BY-NC 4.0 \\
 {\cellcolor{mygrey}}Coswara\cite{bhattacharya2023coswara}   &   IISc    &      \url{https://github.com/iiscleap/Coswara-Data}  & CC BY 4.0  \\
 {\cellcolor{mygrey}}KAUH\cite{fraiwan2021dataset}           &   KAUH   &  \url{https://data.mendeley.com/datasets/jwyy9np4gv/3}   &  CC BY 4.0   \\
 {\cellcolor{mygrey}}Respiratory@TR\cite{altan2017multimedia}&  ITU    & \url{https://data.mendeley.com/datasets/p9z4h98s6j/1}     & CC BY 4.0    \\
 {\cellcolor{mygrey}}SSBPR\cite{xiao2023snoring}             &   WHU   &  \url{https://github.com/xiaoli1996/SSBPR}   &   CC BY 4.0  \\
 {\cellcolor{mygrey}}MMlung\cite{mosuilymmlung}              & UoS     & \url{https://github.com/MohammedMosuily/mmlung}     &Custom license  \\
 {\cellcolor{mygrey}}NoseMic\cite{butkow2024evaluation}      & UoC     &  \url{https://github.com/evelyn0414/OPERA/tree/main/datasets/nosemic}   &Custom license  \\
\bottomrule
\end{tabular}
\end{adjustbox}
\label{tab:data_url}
\end{table}

 \begin{figure*}[t]
    \centering
            \includegraphics[width=0.95\textwidth]{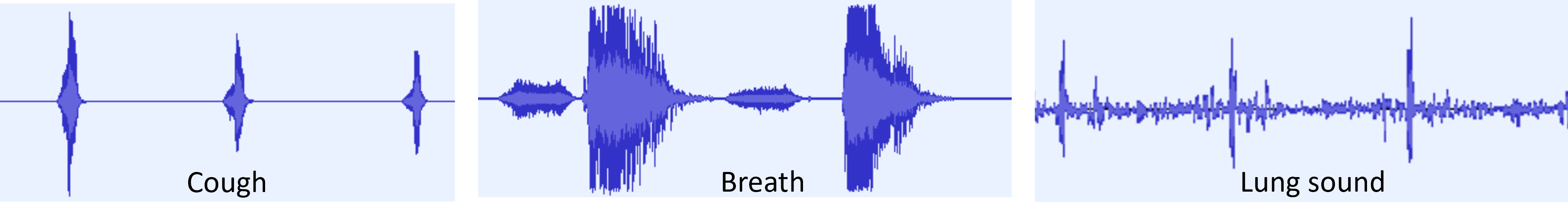}
     \setlength{\abovecaptionskip}{0.1cm}
    \caption{Examples of different respiratory audio modalities used.}
    \label{fig:audiosmaple}
\end{figure*}

\textbf{HF Lung~\cite{hsu2022progressively} }.
HF Lung V2 dataset comprises of HF Lung V1 and HF Lung V1 IP:
The lung sound recordings of HF Lung V1 come from two sources. The first source was a database used in a datathon in Taiwan Smart Emergency and Critical Care (TSECC), 2020, under the license of Creative Commons Attribution 4.0 (CC BY 4.0), provided by the Taiwan Society of Emergency and Critical Care Medicine (TSECCM). Lung sound recordings in the TSECC database were acquired from 261 patients.
The second source was sound recordings acquired from 18 residents of a respiratory care ward (RCW) or a respiratory care center (RCC) in Northern Taiwan between August 2018 and October 2019. The recordings were approved by the Research Ethics Review Committee of Far Eastern Memorial Hospital (case number: 107052-F). Written informed consent was obtained from the 18 patients.

The lung sound recordings of HF Lung V1 IP come from two sources. The Lung sound recordings from the first source are provided by Taiwan Society of Emergency and Critical Care Medicine (TSECCM) acquired from 32 patients by using a commercial digital stethoscope Littmann 3200 (3M). The lung sound recordings of the second source are acquired by from 7 residents of a respiratory care ward (RCW) or a respiratory care center (RCC) in Northern Taiwan between August 2019 and December 2019. The recordings were approved by the Research Ethics Review Committee of Far Eastern Memorial Hospital (case number: 107052-F). Written informed consent was obtained from the 7 patients or their statutory agents.

\textbf{Coswara~\cite{bhattacharya2023coswara}}. The Coswara dataset contains respiratory sounds recorded between April 2020 and February 2022 from 2635 individuals (1819 SARS- CoV-2 negative, 674 positive, and 142 recovered subjects). The respiratory sounds contained nine sound categories associated with variants of breathing, cough and speech. The metadata contains demographic information associated with age, gender and geographic location, as well as the health information relating to the symptoms, pre-existing respiratory ailments, comorbidity and SaRS-CoV-2 test status. 

The data collection procedure was approved by the Institutional Human Ethics Committee, at the Indian Institute of Science, Bangalore. The informed consent was obtained from all participants who uploaded their data records. All the data collected was anonymized and excluded any participant identity information.

\textbf{KAUH~\cite{fraiwan2021dataset}}. 
The KAUH dataset includes sounds from seven ailments (i.e., asthma, heart failure, pneumonia, bronchitis, pleural effusion, lung fibrosis, and chronic obstructive pulmonary disease (COPD) as well as normal breathing sounds. The dataset contains the audio recordings from the examination of the chest wall at various vantage points using an electronic stethoscope. The stethoscope placement on the subject was determined by the specialist physician performing the diagnosis. Each recording was replicated three times corresponding to various frequency filters that emphasize certain bodily sounds. The dataset can be used for the development of automated methods that detect pulmonary diseases from lung sounds or identify the correct type of lung sound.

All study participants (or their parents in the case of underage subjects) provided written informed consent to be included in the study and allowed their data to be shared. This study was approved by the institutional review board at King Abdullah University Hospital and Jordan University of Science and Technology, Jordan (Ref. 91/136/2020). The data collection was carried out under the relevant guidelines and regulations. The authors have the right to share the data publicly.

\textbf{Respiratory@TR~\cite{altan2017multimedia}}.
Respiratory@TR contains lung sounds recorded  from left and right sides of posterior and anterior chest wall and back using two digital stethoscopes in Antakya State Hospital. The chest X-rays and the pulmonary function test variables and spirometric curves, the St. George respiratory questionnaire (SGRQ-C) are collected as multimedia and clinical functional analysis variables of the patients.
The 12 channels of lung sounds are focused on upper lung, middle lung, lower lung and costophrenic angle areas of posterior and anterior sides of the chest. The recordings are validated and labeled by two pulmonologists evaluating the collected chest X-ray, PFT and auscultation sounds of the subjects. Labels fall into 5 COPD severities (COPD0, COPD1, COPD2, COPD3, COPD4). The dataset was released by Iskenderun Technical University, Turkey. Voluntary admittance was evaluated on a voluntary basis form with minimal information. The patients aged 38 to 68 are selected from different occupational groups, socio-economic status and genders for an accomplished analysis of the disorders.

\textbf{SSBPR ~\cite{xiao2023snoring} }.
SSBPR is a snore-based sleep body position recognition dataset consisting of 7570 snoring recordings, which comprises six distinct labels for sleep body position: supine, supine but left lateral head, supine but right lateral head, left-side lying, right-side lying and prone. One of the labels is only present in a few subjects and thus is excluded from the task following the 5-class setup in \cite{xiao2023snoring}.

The data were collected from 20 adult patients who underwent overnight PSG at a local Sleep Medicine Research Center within the hospital. The study was conducted with the approval of the local medical ethics committee, and patients provided signed consent for their participation, including audio and video recordings during sleep. The personal information of the study subjects was collected and stored anonymously to ensure privacy protection.

\textbf{MMLung~\cite{mosuilymmlung} }.
This data was collected from 40 participants (20 male, 20 female) with an age range of 18-85 years old. All participants are English speakers from the UK. Among them, 12 were healthy participants, while the others consisted of seven self-reported COPD patients, seven self-reported asthma patients, and 14 people with other long-term conditions. Ethics approval for this study was obtained from the University of Southampton.

Three devices were used to collect the data: Google Pixel 6 Smartphone with an app installed for the data collection, and an Easy on-PC ultrasonic spirometer by ndd Medical Technologies.   The audio data collection from smartphones was conducted in stereo mode at a sampling rate of 44100 Hz. The data was saved in the \textit{WAV} format. The collection took place in a silent room conditions. The process consisted of collecting data for four audio modalities i.e. cough, vowels, mobile spirometry, and speech via a series of tasks from each participant in a single
session. In this paper, we only include the deep breath and the vowel sound of `o'. 
Ground truth data were collected using a medical-grade spirometer by a healthcare professional as per European Respiratory Society (ATS/ERS) clinical standards. 
However, it should be noted that with any objective measure that is reliant on individual effort, there may always be unforeseen errors (effort dependent blows). This data is available upon request.

\textbf{NoseMic~\cite{butkow2024evaluation} }. 
NoseMic is a subset of the data collected for a respiratory rate estimation project. The audio data was collected using microphones attached close to the nose, and the respiratory dynamics were measured with a Zephyr pressure sensor on the chest. The data was collected in stationary settings, both before and after the participants exercised. A total number of 21 participants were involved, while data from some participants were excluded because of the poor sensing quality.
Audio recordings before and after running were included in our benchmark. Each recording was segmented into 30-second windows with a 15-second overlap. The average respiratory rate of each window was used as the ground truth.

\begin{figure*}[t]
    \centering
    \subfigure[COVID-19 Sounds]{
        \begin{minipage}[b]{0.48\textwidth}
            \centering
            \includegraphics[height = 5.25cm]{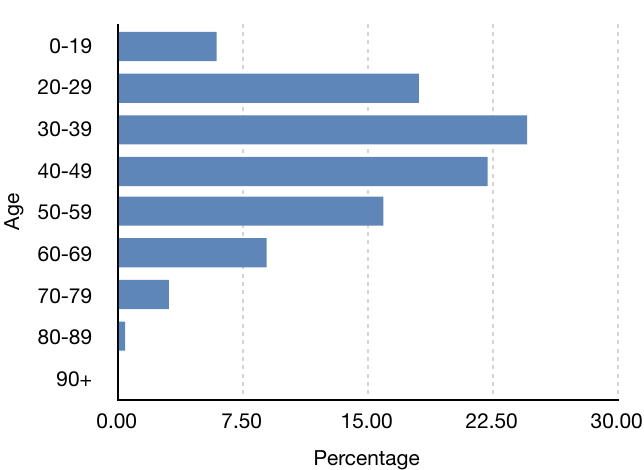}
        \end{minipage}
    }
    \subfigure[UK COVID-19]{
        \begin{minipage}[b]{0.48\textwidth}
            \centering
            \includegraphics[height = 5.25cm]{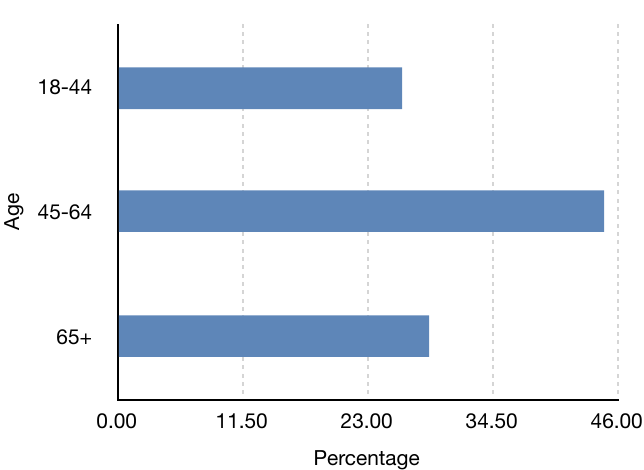}
        \end{minipage}
    }
    \subfigure[COUGHVID]{
        \begin{minipage}[b]{0.48\textwidth}
            \centering
            \includegraphics[height = 5.25cm]{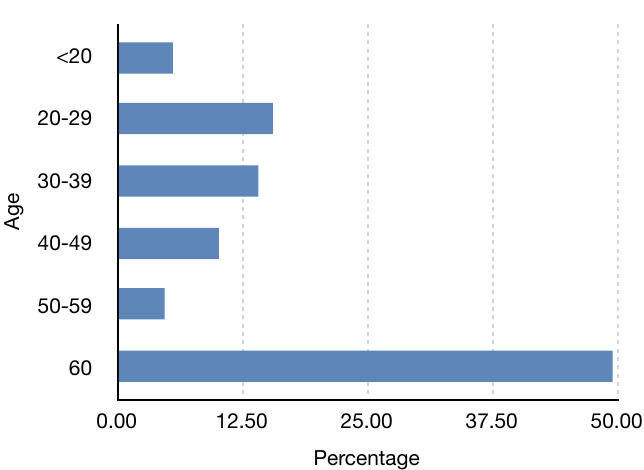}
        \end{minipage}
    }
    \subfigure[ICBHI]{
        \begin{minipage}[b]{0.48\textwidth}
            \centering
            \includegraphics[height = 5.25cm]{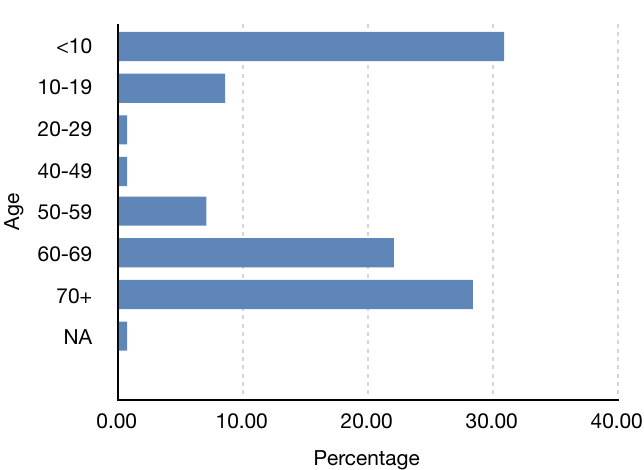}
        \end{minipage}
    }
    \caption{Age distribution of the pretraining datasets.}
    \label{fig:age}
\end{figure*}

\begin{figure*}[t]
    \begin{adjustbox}{width=\textwidth,center}
    \centering
    \subfigure[COVID-19 Sounds]{
        \begin{minipage}[b]{0.24\textwidth}
            \centering
            \includegraphics[height = 4.25cm]{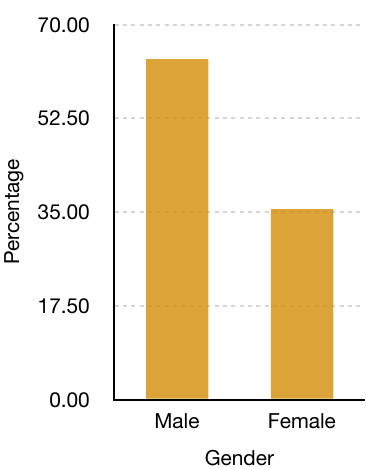}
        \end{minipage}
    }
    \subfigure[COVID-19 UK]{
        \begin{minipage}[b]{0.24\textwidth}
            \centering
            \includegraphics[height = 4.25cm]{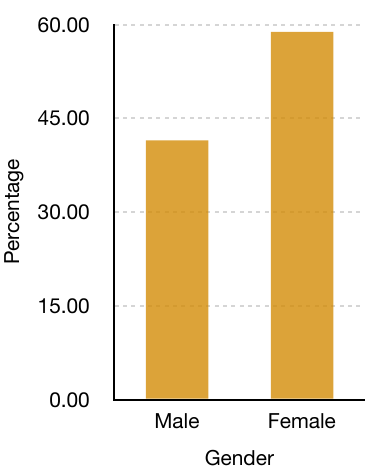}
        \end{minipage}
    }
    \subfigure[COUGHVID]{
        \begin{minipage}[b]{0.24\textwidth}
            \centering
            \includegraphics[height = 4.25cm]{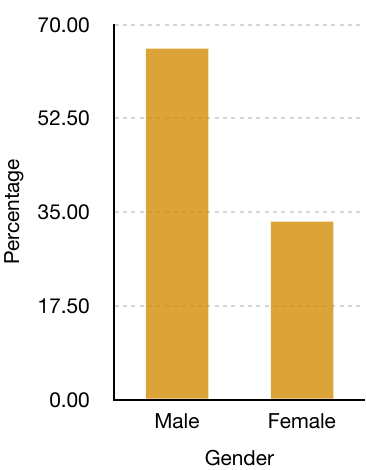}
        \end{minipage}
    }
    \subfigure[ICBHI]{
        \begin{minipage}[b]{0.24\textwidth}
            \centering
            \includegraphics[height = 4.25cm]{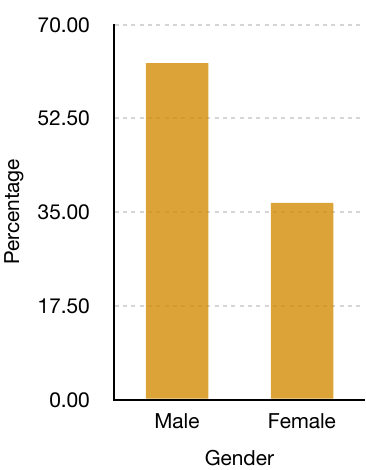}
        \end{minipage}
    }
    \end{adjustbox}
    \caption{Gender distribution of the pretraining datasets.}
    \label{fig:sex}
\end{figure*}

\subsubsection{Pretraining Data Demographics}

Diversity and representativeness of the training data are important for a generalizable model. We examine the demographic distribution of the five datasets used for model pretraining.  The bar plots in \cref{fig:age} and \cref{fig:sex} illustrate the age and gender distributions across four of these datasets. While the demographic details of HF Lung are not publicly available, the data includes 35 male and 21 female subjects, with an average age of 66.58 (according to the paper~\cite{hsu2022progressively}).

\reb{Among the five datasets, COVID-19 Sounds and CoughVID were collected globally, while UK COVID-19 and ICBHI were primarily collected in European countries, and HF Lung was collected in Asian regions. Therefore, our curated data presents a comprehensive geo-distribution, covering participants from different ethnic backgrounds and speaking various languages.}

\begin{figure}[t]
    \begin{adjustbox}{width=\textwidth,center}
        \centering
        \includegraphics{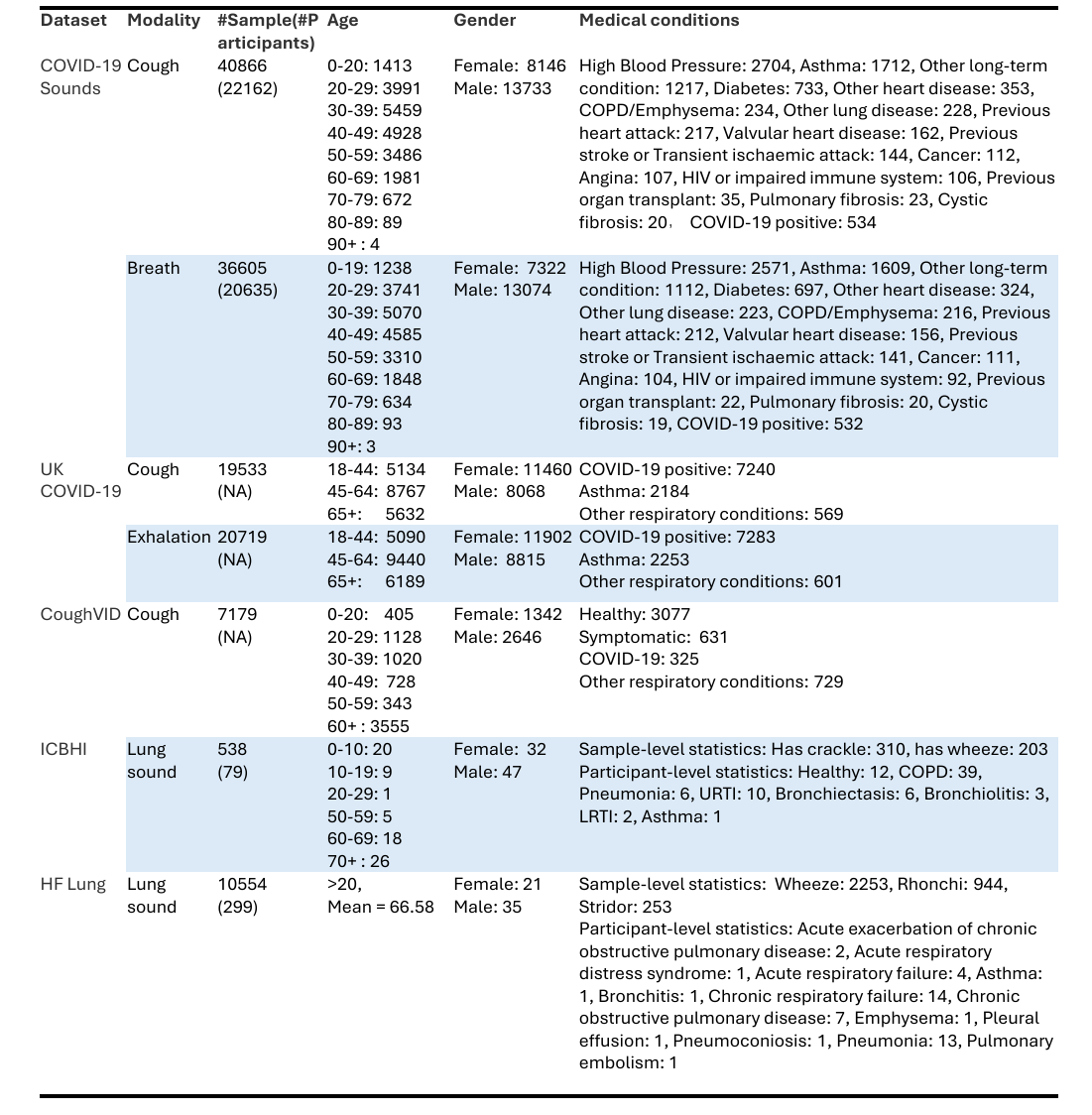}
    \end{adjustbox}
    \caption{Statistics of demographics and medical conditions for datasets used for pretraining.}
    \label{fig:trainingdemo}
\end{figure}

\reb{\cref{fig:trainingdemo} summarizes in detail all demographics and medical conditions for the five datasets used for model pre-training. The five datasets used cover a wide range of respiratory medical conditions. COVID-19 Sounds, UK COVID-19, and CoughVID were collected during the pandemic and include some participants who tested positive or negative for COVID-19. Some of the participants had other conditions such as asthma, COPD, pulmonary fibrosis, cancer, etc. The ICBHI and HF Lung datasets include participants who were either healthy or had various respiratory diseases including asthma, COPD, URTI, Pneumonia, etc. Recordings feature both healthy individuals and those with symptoms such as wheeze, crackles, or rhonchi.}

By integrating these diverse datasets in OPERA, we achieve a more representative and unbiased demographic distribution compared to any single data source. This highlights the importance of uniting varied sources for pretraining a foundational model: not only increasing the number of data samples but also ensuring a more comprehensive distribution.











\subsubsection{Downstream Task Description}

\begin{figure}[t]
    \begin{adjustbox}{width=\textwidth,center}
        \centering
        \includegraphics{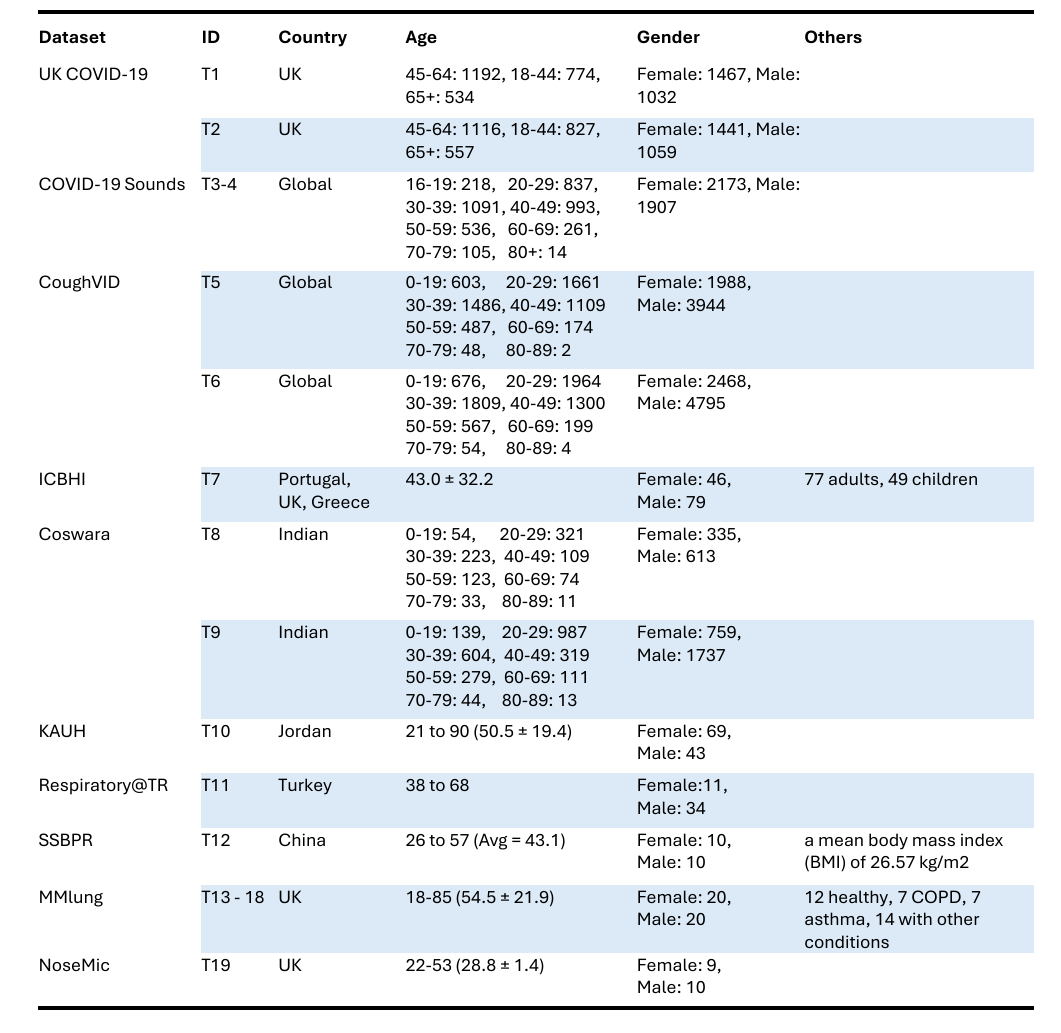}
    \end{adjustbox}
    \caption{Statistics of demographics for downstream tasks.}
    \label{fig:testingdemo}
\end{figure}

Here we give a detailed description of all 19 tasks formulated in the OPERA benchmark. \reb{The demographic statistics are summarized in \cref{fig:testingdemo}.}
The tasks are categorized into three types:
\begin{itemize}[topsep=-1pt, itemsep=0pt, leftmargin=15pt]
    \item \textbf{Binary Classification (Tasks 1-10)}: Tasks requiring prediction of a binary outcome (positive/negative, smoker/non-smoker, etc.) based on respiratory audio recordings.
    \item \textbf{Multi-Class Classification (Tasks 11, 12)}: Tasks involving classification of respiratory audio recordings into one of several predefined categories (5 classes of COPD severity, sleeping position)
    \item \textbf{Regression (Tasks 13-19)}: Tasks aiming to predict continuous values (lung function metrics, respiratory rate) from respiratory audio data.
\end{itemize}

\textbf{Task 1}. Each of the audio in UK COVID-19~\cite{coppock2024audio} has a binary label indicating the COVID-19 test result of the participant. This task is to predict whether the test result is positive based on the exhalation recording, consisting of three successive “ha” exhalation sounds.

\textbf{Task 2}. The data source and prediction target is the same as Task 1, while Task 2 is based on the cough recording consisting of three successive volitional coughs.

\textbf{Task 3}. The audio samples in COVID-19 Sounds~\cite{xia2021covid}  have the reported symptoms at the moment of participation.
This task aims at predicting respiratory abnormalities, where the symptomatic group consists of participants who reported any respiratory symptoms, including dry cough, wet cough, fever, sore throat, shortness of breath, runny nose, headache, dizziness, and chest tightness, while asymptomatic controls are those who reported no symptoms.
The audio data consists of 3 to 5 deep breathing sounds. This task follows the subset and split from~\cite{xia2021covid}, with the training set downsampled.

\textbf{Task 4}. The dataset and prediction target is the same as Task 3, but the audio includes three coughs.

\textbf{Task 5}. Each of the audio in CoughVID\cite{orlandic2021coughvid} contains a cough and is associated with labels of self-reported demographics and COVID-19 status. This task involves predicting the COVID-19 status based on the cough recording.

\textbf{Task 6}. The dataset and audio modality are the same as Task 5, while the prediction target is gender as reported in demographics.

\textbf{Task 7}. The ICBHI~\cite{rocha2019open} dataset contains labels of the diagnosis of the subjects. We use the subset of COPD patients and healthy controls to formulate a binary classification of COPD detection.

\textbf{Task 8}. Each audio in the Coswara~\cite{bhattacharya2023coswara}
 dataset contains a binary label of smoker in the metadata. This task aims to predict the smoker from non-smokers from the cough-shallow audio modality in the dataset, aligning with the implementation in \cite{baur2024hear}.

 \textbf{Task 9}. Each audio in the Coswara~\cite{bhattacharya2023coswara}
 dataset contains a label of sex in the metadata. This task aims to predict this label from the cough-shallow audio modality in the dataset, aligning with the implementation in \cite{baur2024hear}.

\textbf{Task 10}. The KAUH~\cite{fraiwan2021dataset} dataset contains the disease diagnosis labels of the participants. This task aims to use lung sound audio to distinguish patients with COPD and asthma (obstructive lung diseases) from healthy controls.

\textbf{Task 11}. The  Respiratory@TR~\cite{altan2017multimedia} dataset associates each audio with a COPD severity label from 0 to 4. This task aims to predict this severity level from lung sounds.

\textbf{Task 12}. The SSBPR~\cite{xiao2023snoring} dataset associates each snoring audio with a label of the body position: supine, supine but left lateral head, supine but right lateral head, left-side lying, right-side lying and prone. The last class is excluded here as it is only present in some of the male participants. Thus this task aims to predict one of the five body positions from the snoring sounds.

\textbf{Task 13}. Spirometry is a gold standard for diagnosing Long-term respiratory illnesses like COPD and Asthma.
It is a lung health test that requires specialized equipment and trained healthcare experts, making it expensive and difficult to scale. Moreover, blowing into a spirometer can be quite hard for people suffering from pulmonary illnesses. To address this problem, researchers aim to develop audio-based testing methods without requiring the best efforts from patients. MMLung~\cite{mosuilymmlung} was collected for this purpose. Task 13 evaluates how accurate the forced vital capacity (FCV) can be estimated from a deep breath sound. 

\textbf{Task 14}. Similar with Task 13 , Task 14 evaluates how accurate the forced expiratory volume in 1 second (FEV1) can be estimated from a deep breath sound.

\textbf{Task 15}. While FEV1 and FVC are very personal, the ratio between them is the proportion of lung capacity that can be exhaled in the first second. It is expressed as a percentage and is used to diagnose and determine the severity of obstructive and restrictive lung diseases. Task 15 uses breathing sounds to estimate this ratio.

 \textbf{Task 16}. Task 16 again aims to evaluate an individual's FVC, similar to Task 13. However, a vowel sound is used, i.e., the participant speaks out the `o' sound for as long as possible.

 \textbf{Task 17}. Task 17 involves the use of  `o' vowel sound for FEV1 estimation. 

\textbf{Task 18}. This task predicts the ratio between FEV1 and FVC from the collected `o' vowel sounds. 

 \textbf{Task 19}.  Continuous respiratory rate (RR) monitoring is integral to mobile healthcare and fitness tracking, offering valuable insights into longitudinal health and wellness due to its strong correlations with both physical and mental health. This task involves the estimation of RR from 30 seconds of breathing sounds.

\subsection{Implementation Details}
\label{app:imple}


All of the experiments are implemented in Python 3.10.4, with main supporting libraries: PyTorch, Librosa, PyTorch Lightning, numpy, with the exact environment detailed in `environment.yml' in the code repository.
All our experiments are conducted using a NVIDIA A100 GPU with 80GB memory. Our code is accessible from \url{https://github.com/evelyn0414/OPERA}.

 \subsubsection{Pretraining Models and Methods}

We pre-train our models on a combination of seven sets of data derived from the first five data sources in \cref{tab:license} (including separate modalities from COVID-19 Sounds and UK COVID-19). Each set of data is split into batches of equal length to ensure consistent data processing. These batches maintain both modality and source homogeneity.
We then randomly shuffle the batches and reserve 10\% for validation. Due to inherent variations in audio length within individual batches, we employ random cropping of spectrograms. Crop lengths for each of the seven datasets are detailed in \cref{tab:data}, and the crop methods depend on the pretraining methods, which will be elaborated on below.
Two representative SSL approaches are adopted: contrastive learning-based methods and generative pretraining-based methods, to pretrain three models. The high-level reasoning behind this is that if an encoder can distinguish the source of audio segments (contrastive) or reconstruct masked spectrograms (generative), it is expected to encode useful and generalizable acoustic features. Specifically:

\begin{figure}[t]
    \begin{adjustbox}{width=\textwidth,center}
        \centering
        \includegraphics{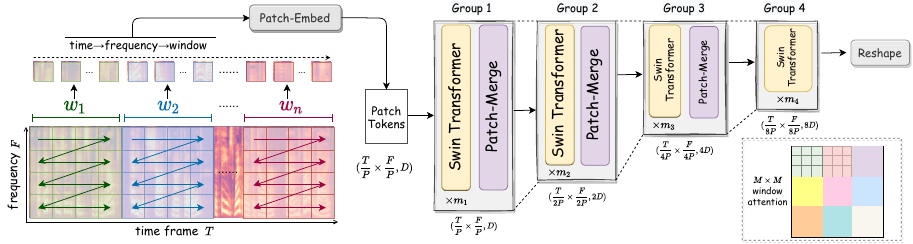}
    \end{adjustbox}
    \caption{The hierarchical token-semantic audio transformer architecture, from \cite{chen2022hts}.}
    \label{fig:htsat}
\end{figure}
    
\textbf{OPERA-CT}: 
OPERA-CT is a contrastive learning-based transformer model. Following~\cite{saeed2021contrastive}, we randomly crop two segments from a spectrogram and regard them as a positive pair. Segments from different samples within one batch are regarded as negative pairs. As shown in \cref{fig:pretrain}(a), an encoder network (a transformer here) extracts features from these segments, and a projector (a multi-layer perception) maps them into a low-dimensional representation space, where bilinear similarity is calculated as,
\begin{equation}
    s(x, x') = g(f(x))^T W g(f(x')).
\end{equation}

The optimization objective aims to maximize the similarity between positive pairs and minimize it for negative pairs. The loss function for this instance discrimination objective is a multi-class cross entropy applied to similarities,

\begin{equation}
    \mathcal{L} = - \log \frac{\exp \left( s(x, x^+) \right) }{\sum_{x^- \in \mathcal{X}^-(x) \cup \{x^+\}}\exp \left( s(x, x^-) \right) },
\end{equation}

where $x^+$ is the positive anchor for $x$ and $\mathcal{X}^-(x) $ refers to negative distractors.

Specifically, the transformer we employ is a hierarchical token-semantic audio transformer~\cite{chen2022hts}, which improves the computing and memory efficiency of the typical vision transformer for spectrograms. A patch size of $4 \times 4$ is used and the output feature dimension is 768. The encoder has 31M trainable parameters.

\textbf{OPERA-CE}: Similar to OPERA-CT, CE leverages a contrastive pre-training approach. However, it utilizes a more lightweight and efficient CNN encoder (EfficientNet-B0)~\cite{tan2019efficientnet}. The architecture is detailed in \cref{tab:efficientnet}. This encoder outputs a feature dimension of 1280 and has approximately 4M trainable parameters.

\begin{table}
\centering
\caption{The EfficientNet-B0 architecture.}
\label{tab:efficientnet}
\begin{adjustbox}{width=0.5\textwidth,center}
\begin{tabular}{llll} 
\toprule
Layer                & Kernel Size & \#channels & \#layers  \\
\midrule
Input                &      -       & 32         & 1         \\
MBConv1              & 3x3         & 16         & 1         \\
MBConv6              & 3x3         & 24         & 2         \\
MBConv6              & 5×5         & 40         & 2         \\
MBConv6              & 3x3         & 80         & 3         \\
MBConv6              & 5x5         & 112        & 3         \\
MBConv6              & 5x5         & 192        & 4         \\
MBConv6              & 3x3         & 320        & 1         \\
Conv head \& Avg Pooling  &             & 1280       & 1         \\
\bottomrule
\end{tabular}
\end{adjustbox}
\end{table}

\begin{figure}[t]
    \begin{adjustbox}{width=0.9\textwidth,center}
        \centering
        \includegraphics{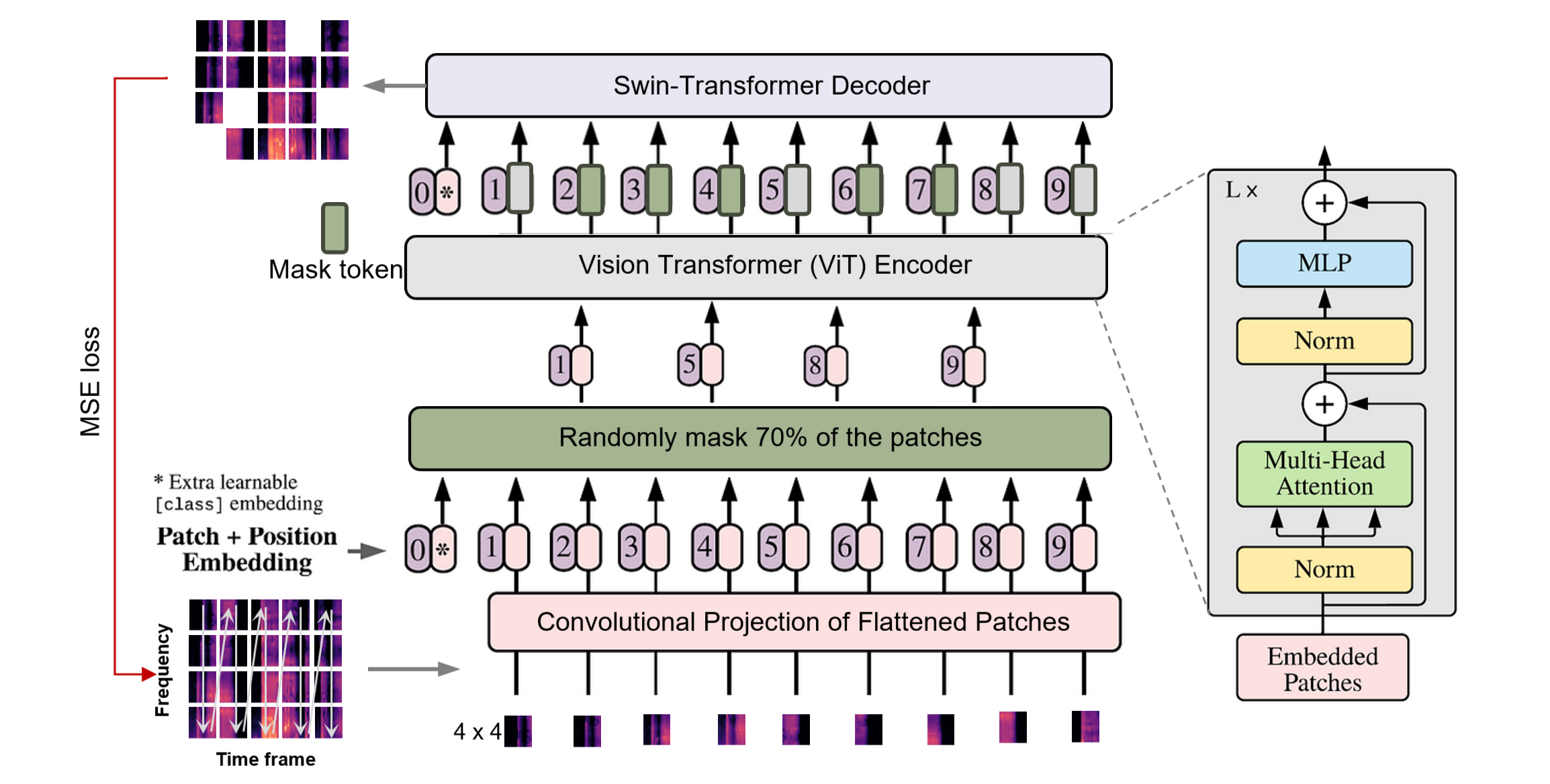}
    \end{adjustbox}
    \caption{OPERA-GT architecture.}
    \label{fig:gt}
\end{figure}

 \textbf{OPERA-GT}: OPERA-GT is a generative pretrained transformer model. It uses a masked auto-encoder to extract useful features from masked spectrograms, which a decoder then uses to reconstruct the original spectrograms, as illustrated in \cref{fig:pretrain}(b). Following~\cite{baade2022mae}, we employ a vision transformer as the encoder (21M trainable parameters) and a lightweight swin-transformer (12M trainable parameters) as the decoder. The detailed architecture is shown in \cref{fig:gt}.

To train this model, spectrograms from each dataset are cropped to equal lengths, as summarized in \cref{tab:data}, and then split into patches of $4 \times 4$. Considering the varying lengths of different modalities, our model uses a unique patching order and accommodates any input length (no larger than the number of positional embeddings), as indicated by the arrows in \cref{fig:gt}. Each patch is converted into a patch embedding via a 2-dimensional convolutional layer with a kernel size of $4 \times 4$ and a channel number of 384. We randomly mask 70\% of patches per spectrogram and only feed the embeddings of the visible patches into the encoder. The encoder is a typical vision transformer with $l=12$ blocks and 2 heads in each block. The output feature dimension is 384.

To reconstruct the spectrograms, both the embeddings of the masked patches and the new embeddings from the encoder are fed into the decoder. The decoder is a typical swin-transformer with both local and global attention. The output of the decoder is an array resembling a spectrogram. Mean square error loss is used for optimization, and only the masked pixels are considered in the loss,
\begin{equation}
    \mathcal{L}_\text{MSE} = \frac{1}{N} \sum_{i=1}^{N} (y_i - \hat{y}_i)^2, 
\end{equation}
where $y$ is the vector only with the masked pixels in the $i$-th spectrogram.

\subsubsection{Benchmark implementation details}


Within our benchmark of downstream tasks, we have four baselines to compare with the OPERA models. Opensmile is chosen as a baseline representing the traditional feature extraction methods.
VGGish, AudioMAE and CLAP are chosen as baselines for this study since they are open-source pretrained models representing the cutting edge of deep learning approaches.

\textbf{Opensmile}. OpenSMILE~\cite{eyben2010opensmile} is a powerful tool for extracting features from audio data. It offers pre-defined feature sets designed to capture various aspects of an audio signal. This established toolkit serves as a strong baseline for traditional feature extraction. It offers a diverse set of handcrafted features, providing a foundation for comparison.

\textbf{VGGish}. The VGGish model~\cite{hershey2017cnn} is a modified VGG model using mel spectrograms as input, pretrained to classify the soundtracks of a dataset of 70M training videos (5.24 million hours) with 30,871 video-level labels. 

\textbf{AudioMAE}. AudioMAE~\cite{huang2022masked} leverages self-supervised learning for audio, inspired by image-based Masked Autoencoders (MAE)~\cite{he2022masked}. During training, AudioMAE masks a high proportion (70\%) of the spectrogram patches and feeds the remaining unmasked tokens through a transformer encoder, which then attempts to reconstruct the original spectrogram. This process forces the model to learn robust features by relying on context and relationships within the spectrogram. 

\textbf{CLAP}. The CLAP  model is trained under natural language supervision, leveraging text descriptions to learn about audio concepts. It utilizes two encoders: one for processing audio spectrograms and another for handling text descriptions. Through a contrastive learning approach, CLAP brings these audio and text features into a shared space and encourages similarity within the same audio-text pair.

\begin{table}[t]
\centering
\caption{Number of parameters and feature dimension of all the models.}
\label{tab:dim}
\begin{adjustbox}{width=\textwidth,center}
\begin{tabular}{lccccccc} 
\toprule
                  & Opensmile & VGGish & AudioMAE & CLAP & \textbf{OPERA-CT} & \textbf{OPERA-CE} & \textbf{OPERA-GT}  \\
\midrule
\# Parameters (M) & -         & 62  & 86     & 80 & 31              & 4              & 21               \\

Input length (s)     & -         & 1     & 10      & 5  & <32              & >1.5              & <8.18              \\

Feature Dim. & 988       & 128    & 768      & 1024 & 768               & 1280              & 384                \\
\bottomrule
\end{tabular}
\end{adjustbox}
\end{table}

For baselines, both the data pre-processing and feature extraction strictly follow their official implementation. For our pretrained models, the same audio preprocessing is used as in pretraining. 
The required audio input length is also summarized in \cref{tab:dim}. 

Our OPERA models can accept audio input of different lengths. Specifically, OPERA-CT has an interpolation step that transforms all spectrogram inputs to the same size, fitting the hierarchical structure of the model~\cite{chen2022hts}. Audio longer than the maximum input length of about 32 seconds will need to be cropped, although this is not relevant to our downstream tasks.
OPERA-CT is a CNN model with a pooling layer, allowing it to always output fixed-length features. However, it requires a minimum length of 1.5 seconds (the input size must be larger than the kernel size). OPERA-GT, a transformer model, incorporates a special patching method (see \cref{fig:gt}) that allows it to accept varying lengths of audio shorter than its maximum input length of 8.18 seconds. For input audio exceeding 8 seconds, we segment the audio into short frames with overlaps, feed them into the model, and use the averaged representation of these frames as the final embedding~\cite{huang2022masked}.



Our evaluation employs linear evaluation for all downstream tasks. This technique leverages the pre-trained model's weights without modification, preserving their learned features. A new linear layer, sized according to the feature dimension (see \cref{tab:dim}) and the number of output classes (or 1 dimension for regression) in the specific downstream task, is added on top of the pre-trained model's output. This approach offers an efficient way to transfer the knowledge of the pre-trained models without extensive fine-tuning of the entire model and can be used for tasks with very limited data size. For classification tasks, a standard cross-entropy loss is used. For regression tasks, an MAE loss is used. A L2 regularization of $10^{-5}$ is employed.







\newpage

\subsection{Pretraining Results}
\label{app:pretrain}

\begin{figure*}[th]
\begin{adjustbox}{width=1.1\textwidth,center}
    \centering
    \subfigure[OPERA-CT]{\includegraphics[height=4.6cm]{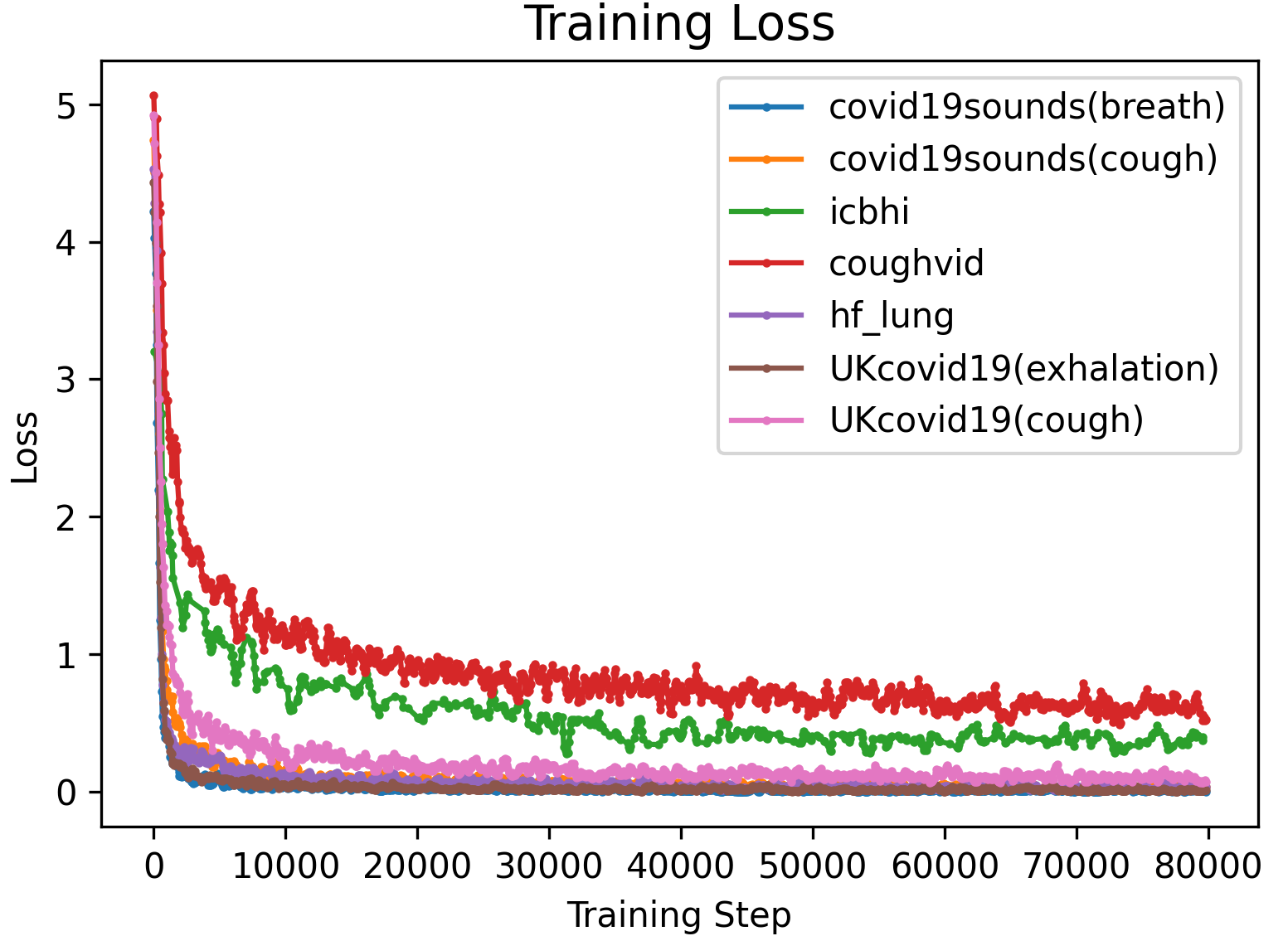} \label{fig:loss:ct}}
    \subfigure[OPERA-CE]{ \includegraphics[height=4.6cm]{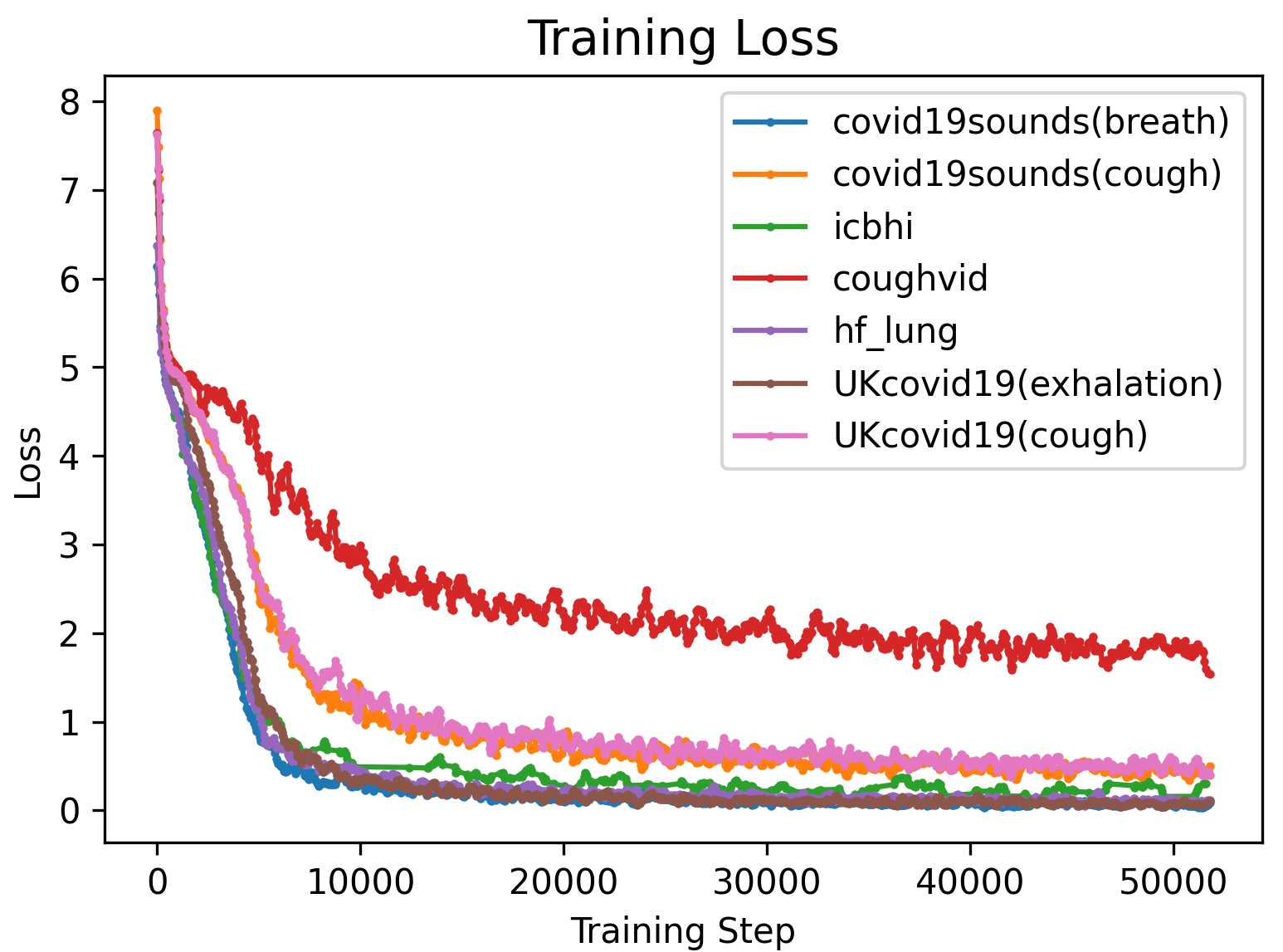} \label{fig:loss:ce}}
    \subfigure[OPERA-GT]{ \includegraphics[height=4.6cm]{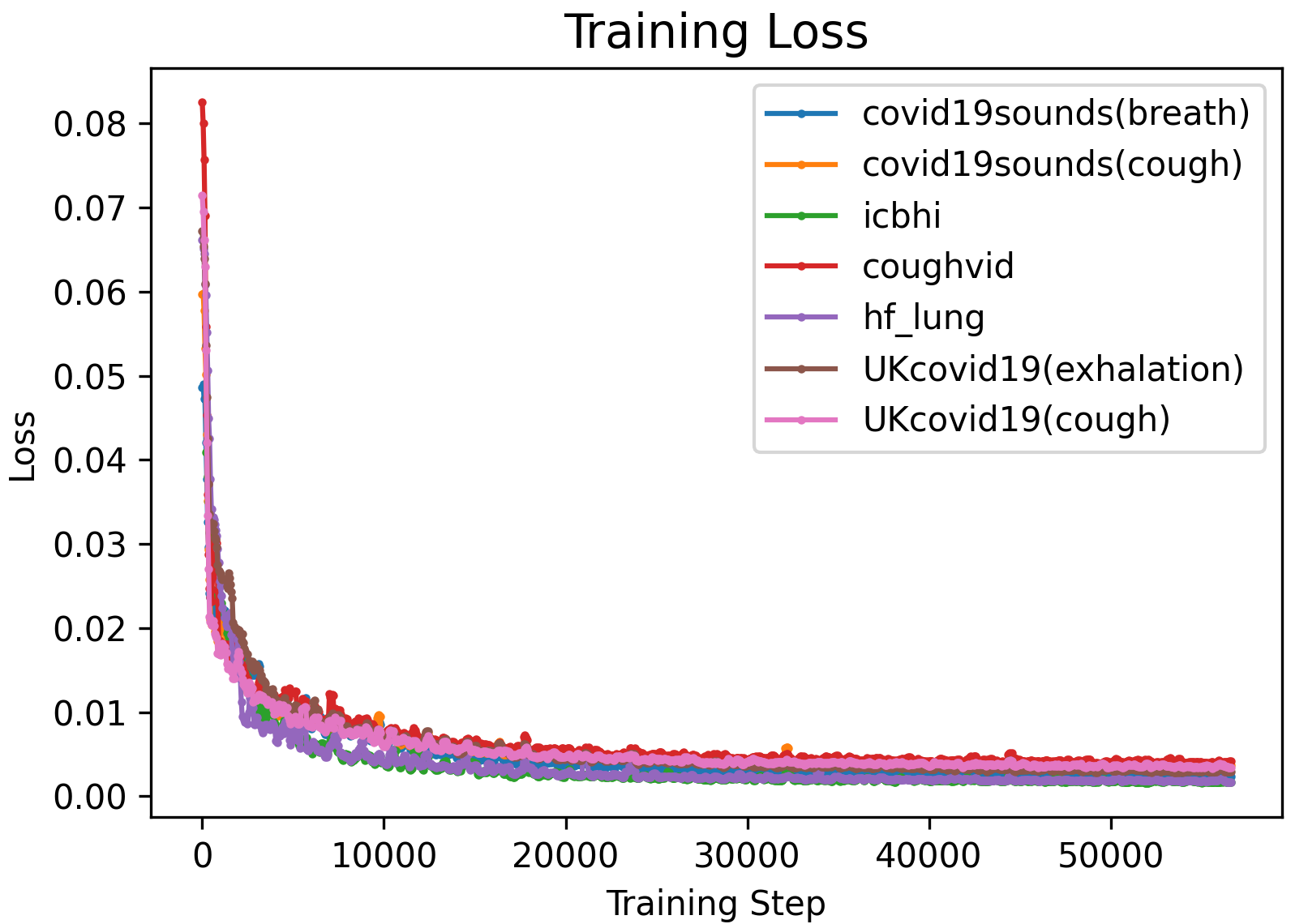}  \label{fig:loss:gt}}
     \setlength{\abovecaptionskip}{0.1cm}
\end{adjustbox}
    \caption{Training loss of the three OPERA models. The OPERA-GT and OPERA-CE use contrastive instance discrimination loss, while OPERA-GT uses generative mean square error loss.}
    \label{fig:loss:training}
\end{figure*}

\begin{figure*}[th]
\begin{adjustbox}{width=1.1\textwidth,center}
    \centering
    \subfigure[OPERA-CT]{\includegraphics[height=4.8cm]{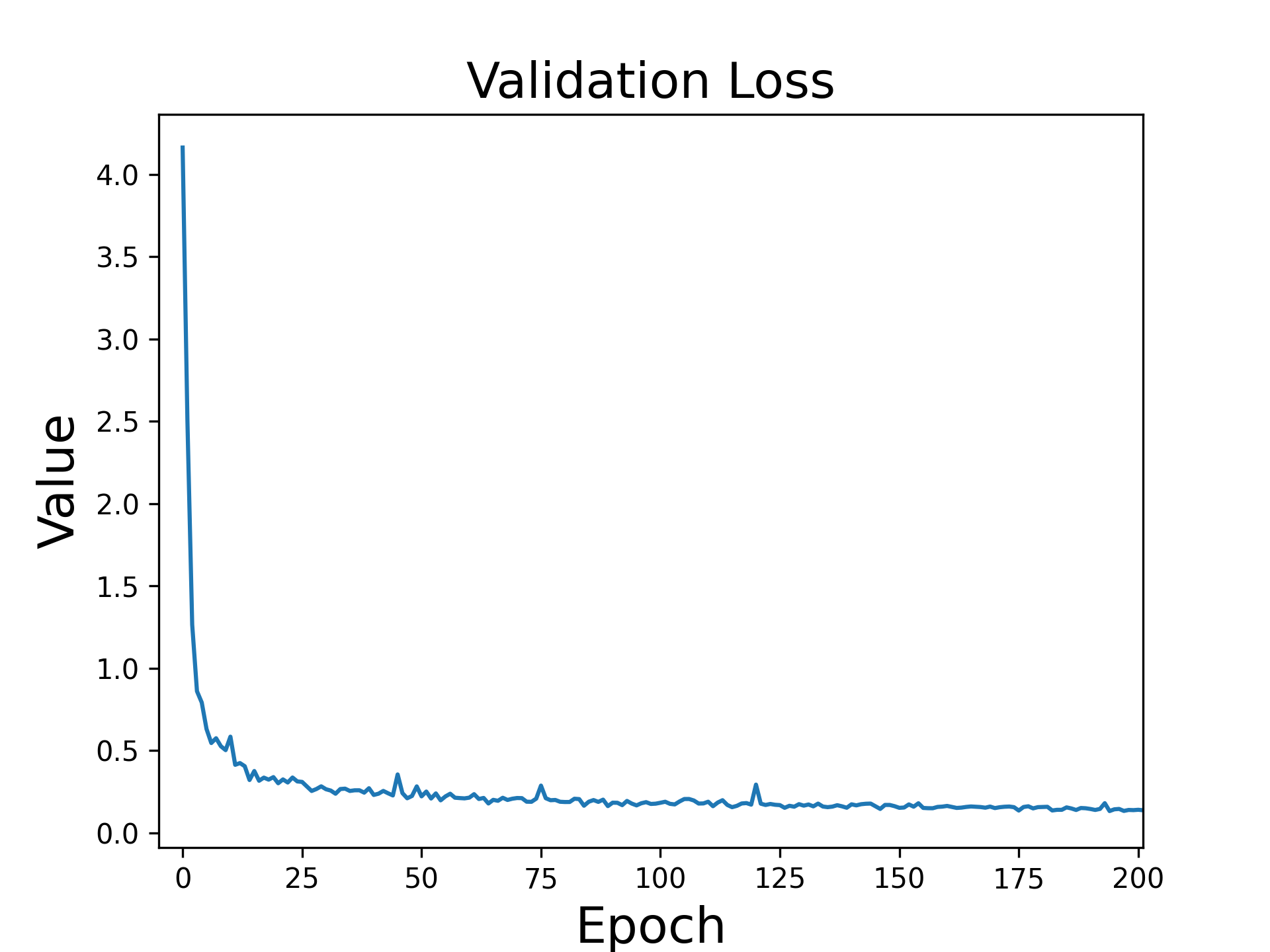} \label{fig:loss:ct}}
    \subfigure[OPERA-CE]{ \includegraphics[height=4.8cm]{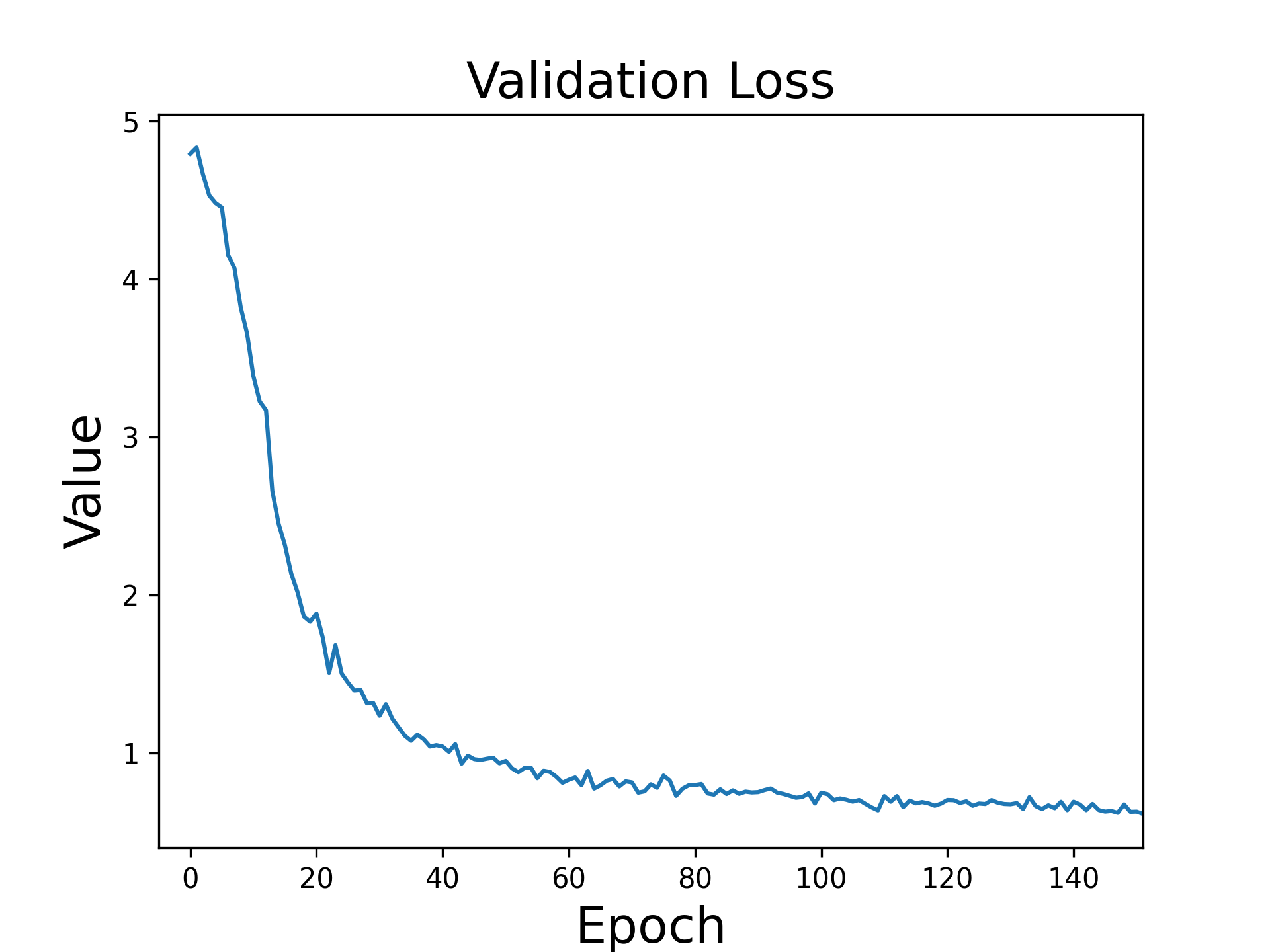} \label{fig:loss:ce}}
    \subfigure[OPERA-GT]{ \includegraphics[height=4.8cm]{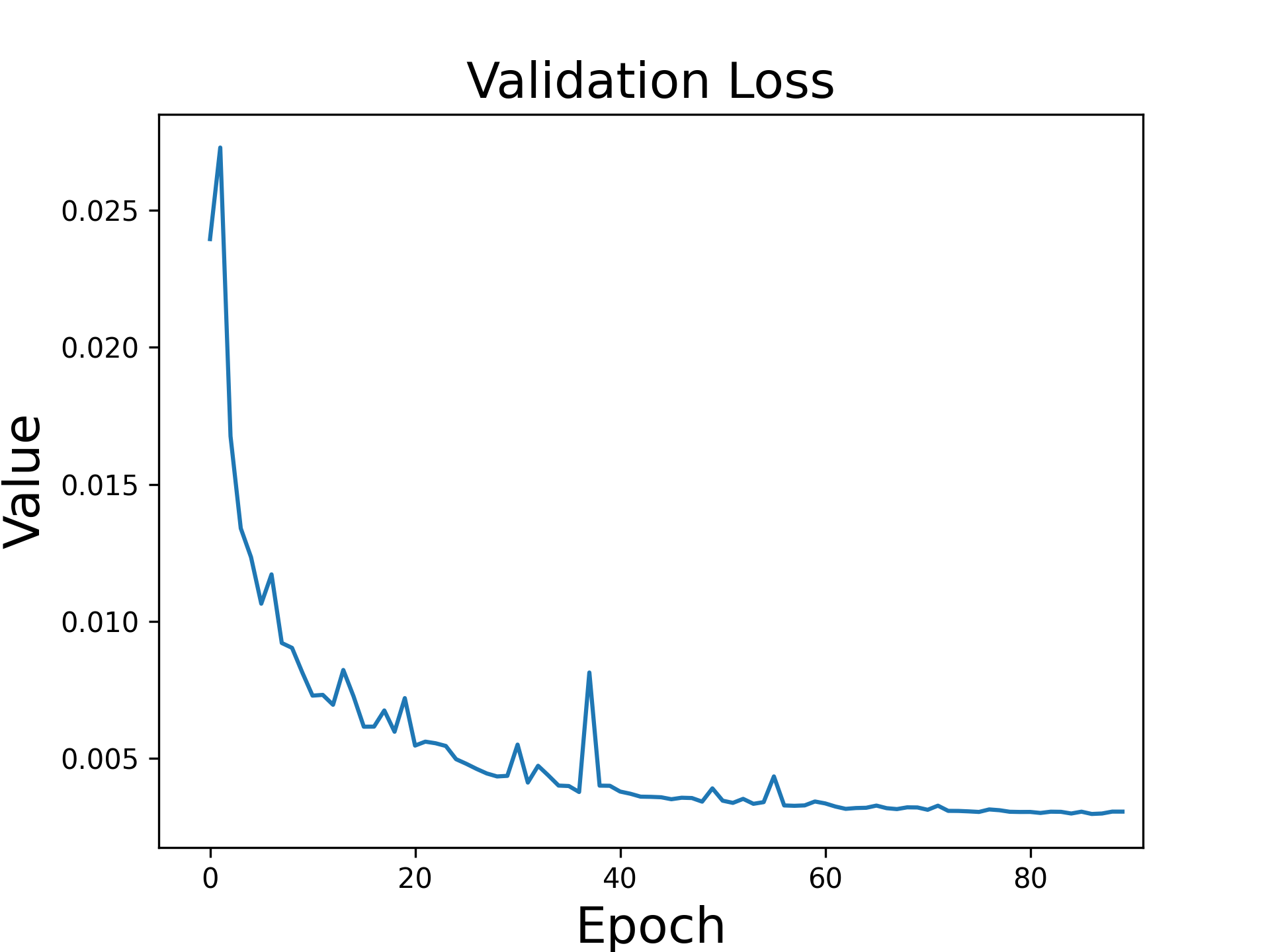}  \label{fig:loss:gt}}
     \setlength{\abovecaptionskip}{0.1cm}
\end{adjustbox}
    \caption{Validation loss of the three OPERA models. The OPERA-GT and OPERA-CE use contrastive instance discrimination loss, while OPERA-GT uses generative mean square error loss.}
    \label{fig:loss}
\end{figure*}

\textbf{Pretraining loss.}  We showcase the training process of our three OPERA models here. Specifically, 
\cref{fig:loss:training} exhibits the training loss of different subsets of the data, converging at different speeds and levels, due to heterogeneity in data quality, data modality, etc. \cref{fig:loss} present the evolution of the loss on the validation set (a set combined a small proportion from all the data resource). It demonstrates a continued decay until convergence.


\begin{figure*}[ht]
\begin{adjustbox}{width=0.95\textwidth,center}
    \centering
    \subfigure[COVID-19 Sounds (breath)]{\includegraphics[height=4.5cm]{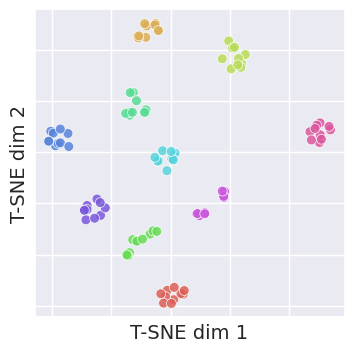} }
    \subfigure[UK COVID-19 (cough)]{\includegraphics[height=4.5cm]{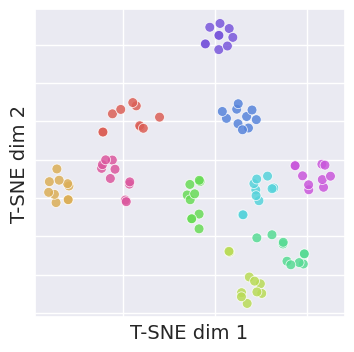}}
    \subfigure[HF Lung (lung sounds)]{\includegraphics[height=4.5cm]{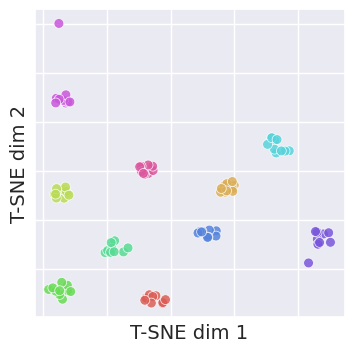} }
     \setlength{\abovecaptionskip}{0.1cm}
\end{adjustbox}
    \caption{T-SNE visualization result of features from OPERA-CT on the held-out validation of pretraining data. Each dot is an audio segment and the same color represents the same audio recording.  It can be seen that audio segments from the same recording are close to each other while far away from other recordings in the embedding space.  }
    \label{fig:tsne:ct}
\end{figure*}
\begin{figure*}[ht]
\begin{adjustbox}{width=0.95\textwidth,center}
    \centering
    \subfigure[COVID-19 Sounds (breath)]{\includegraphics[height=4.5cm]{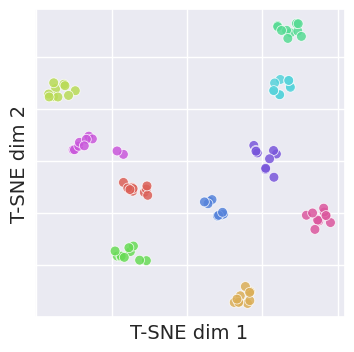}} 
    \subfigure[UK COVID-19 (cough)]{\includegraphics[height=4.5cm]{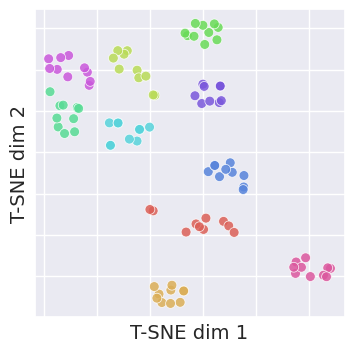}}
    \subfigure[HF Lung (lung sounds)]{\includegraphics[height=4.5cm]{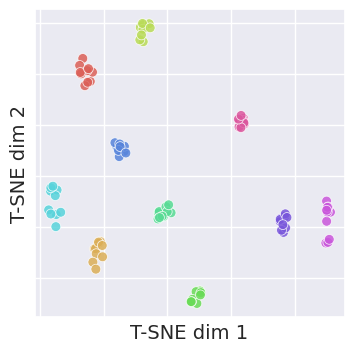} }
     \setlength{\abovecaptionskip}{0.1cm}
\end{adjustbox}
    \caption{T-SNE visualization result of features from OPERA-CE on the validation data.  }
    \label{fig:tsne:ce}
\end{figure*}

\newpage
\textbf{Embedding distribution analysis for constructive pretraining.}
\cref{fig:tsne:ct} and \cref{fig:tsne:ce}
 present the T-SNE visualization applied to features extracted from the contrastive pretraining models on the held-out test set of pretraining data. The visualization depicts four random crops of the same audio sample (the same color) close together in the embedding space. This suggests that the model can effectively capture the underlying characteristics of the audio data despite variations introduced by cropping.

\newpage

\begin{figure*}[th]
\begin{adjustbox}{width=1.2\textwidth,center}
    \centering
    \subfigure[Original spectrogram]{\includegraphics[height=2.8cm]{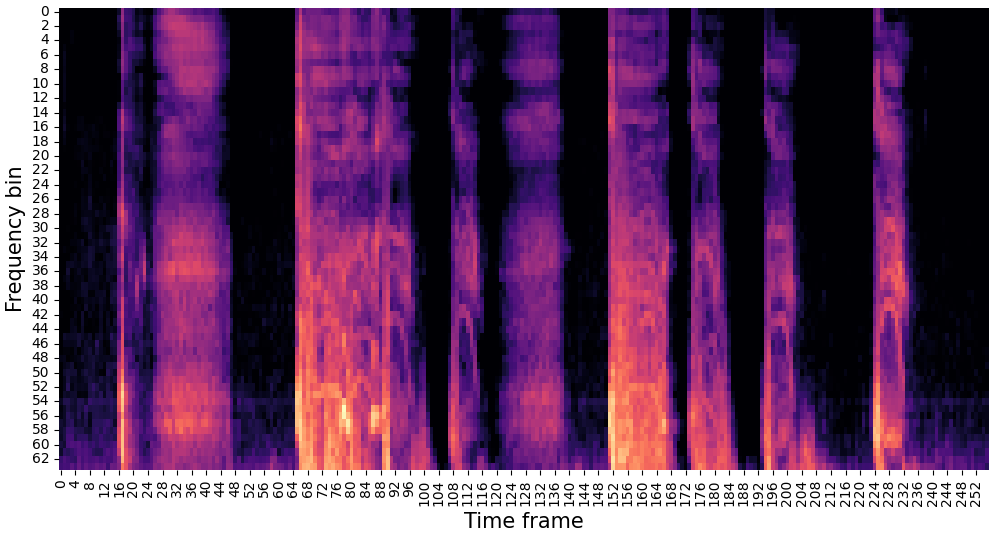}} 
    \subfigure[Masked spectrogram]{\includegraphics[height=2.8cm]{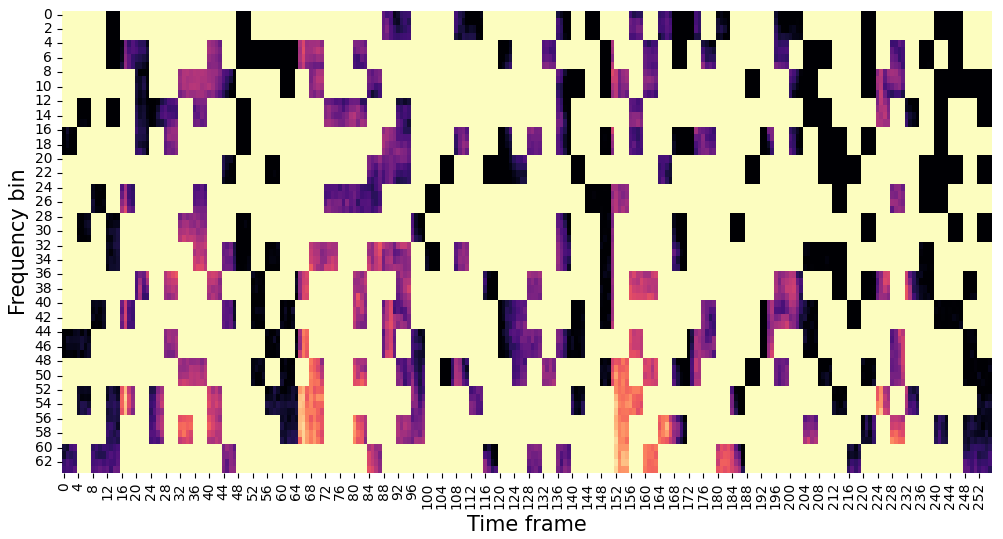}}
    \subfigure[Reconstructed spectrogram]{\includegraphics[height=2.8cm]{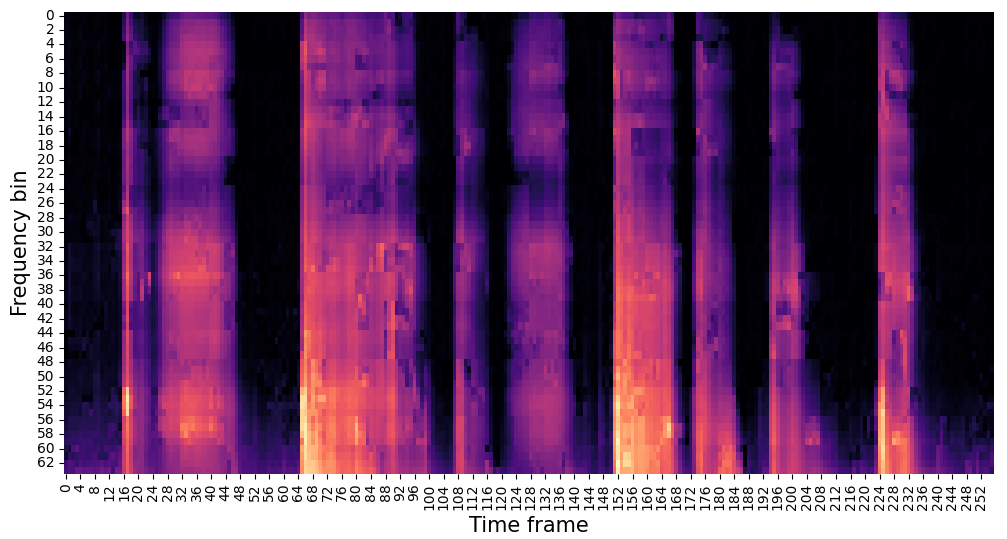} }
     \setlength{\abovecaptionskip}{0.1cm}
\end{adjustbox}
    \caption{Reconstruction result for a breath sound recording (cropped into 8s) from COVID-19 Sounds dataset.   }
    \label{fig:mae:breth}
\end{figure*}
\begin{figure*}[th]
\begin{adjustbox}{width=1.2\textwidth,center}
    \centering
    \subfigure[Original spectrogram]{\includegraphics[height=2.8cm]{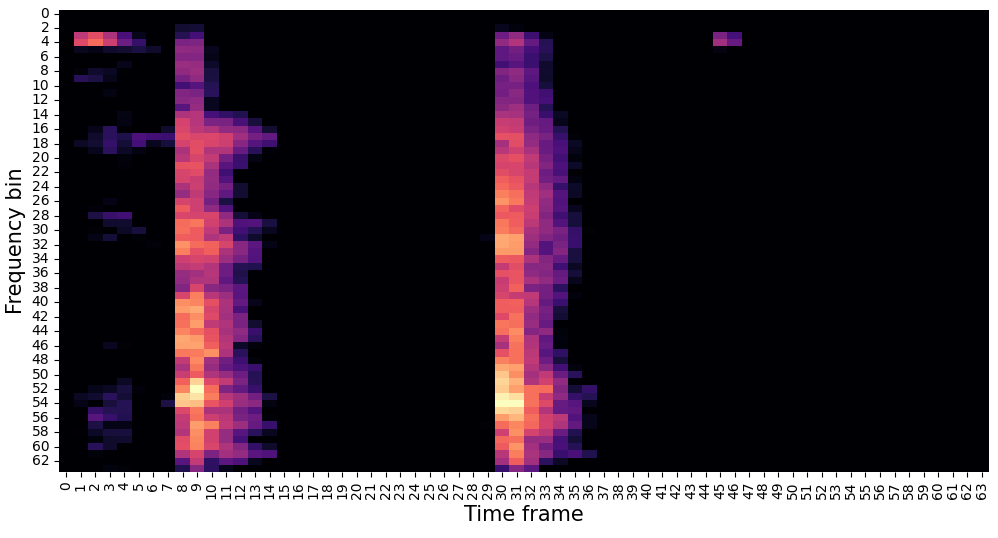}} 
    \subfigure[Masked spectrogram]{\includegraphics[height=2.8cm]{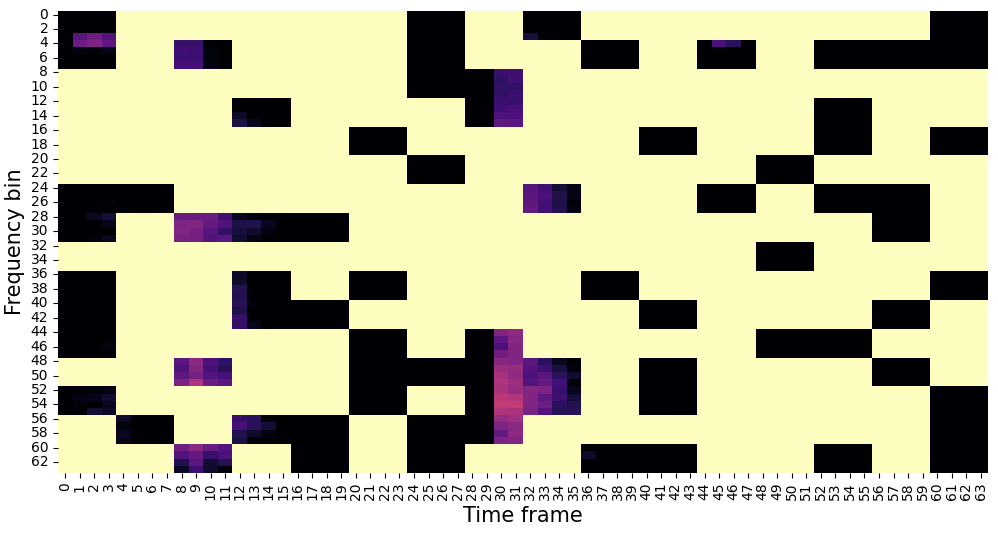}}
    \subfigure[Reconstructed spectrogram]{\includegraphics[height=2.8cm]{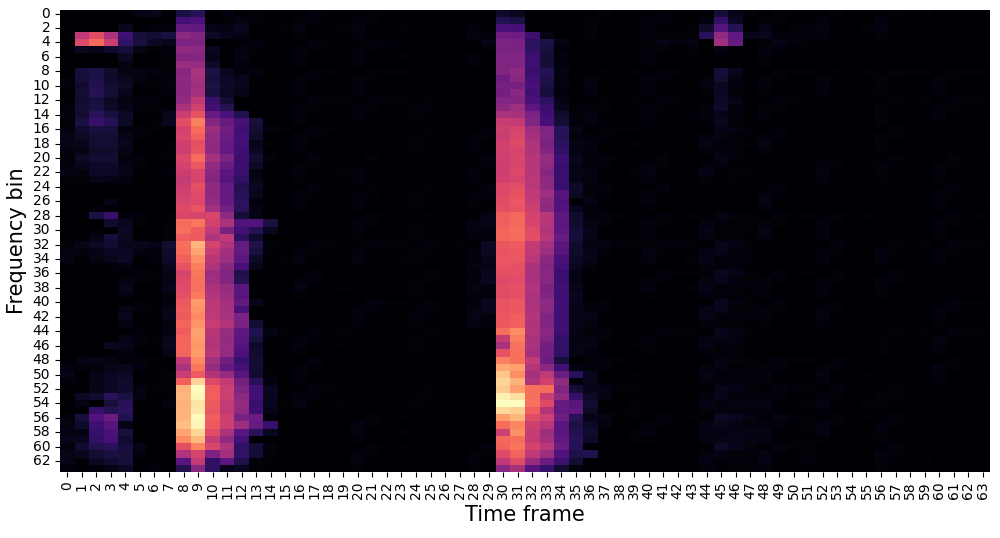} }
     \setlength{\abovecaptionskip}{0.1cm}
\end{adjustbox}
    \caption{Reconstruction result for a cough sound recording (cropped into 2s) from COUGHVID dataset.   }
    \label{fig:mae:cough}
\end{figure*}
\begin{figure*}[th]
\begin{adjustbox}{width=1.2\textwidth,center}
    \centering
    \subfigure[Original spectrogram]{\includegraphics[height=2.8cm]{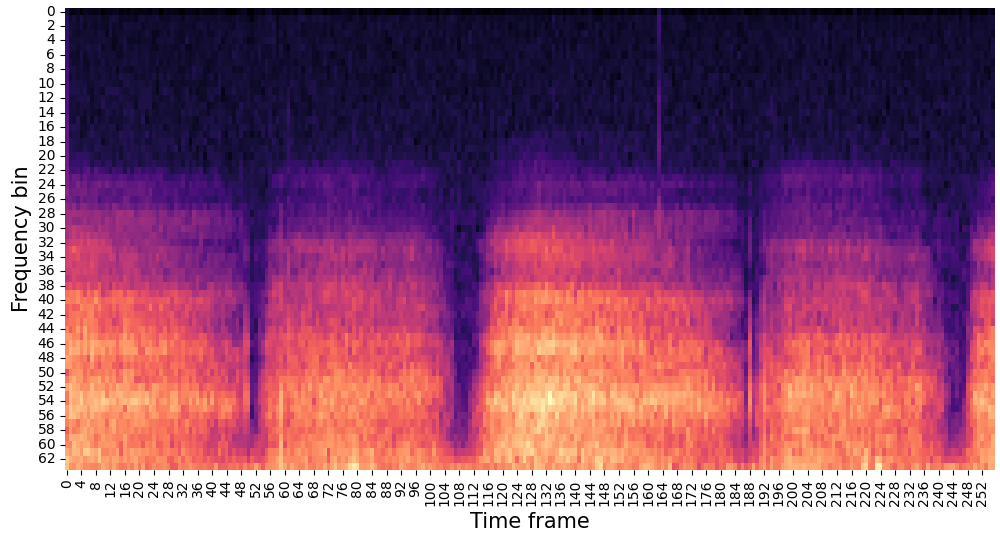}} 
    \subfigure[Masked spectrogram]{\includegraphics[height=2.8cm]{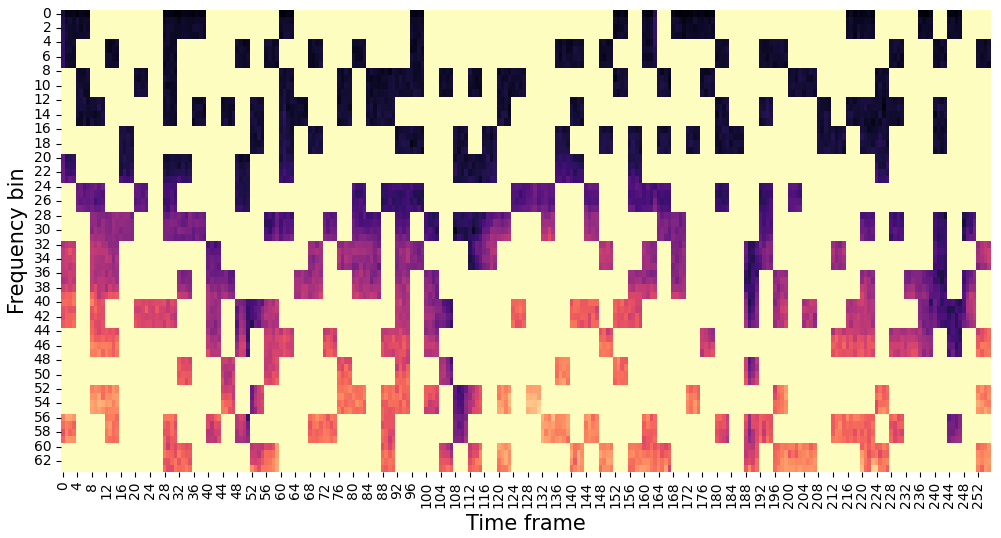}}
    \subfigure[Reconstructed spectrogram]{\includegraphics[height=2.8cm]{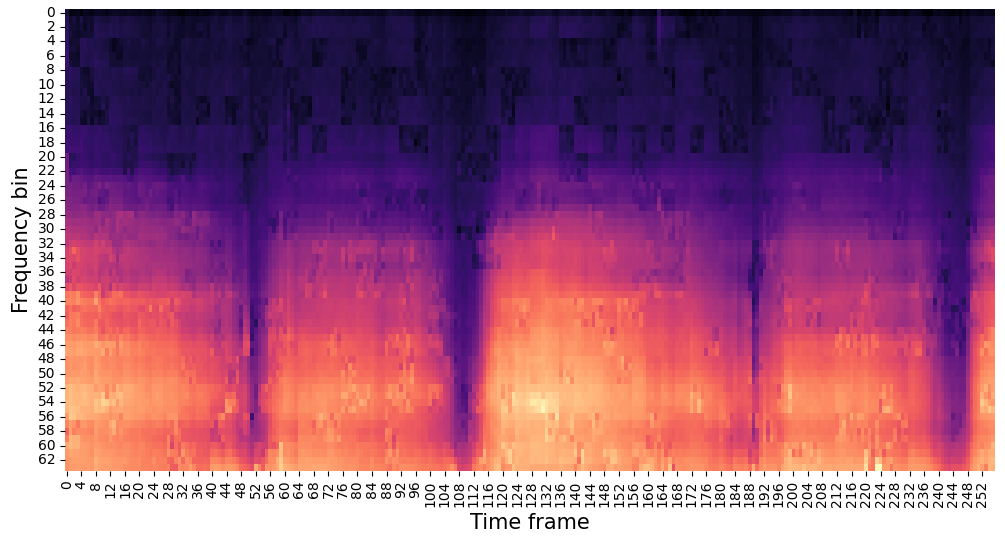} }
     \setlength{\abovecaptionskip}{0.1cm}
\end{adjustbox}
    \caption{Reconstruction result for a lung sound recording (cropped into 8s) from ICBHI dataset.   }
    \label{fig:mae:lung}
\end{figure*}

 \textbf{Spectrogram reconstruction result for generative pretraining}.  OPERA-GT aims to learn a useful encoder by extracting features that can be used to reconstruct the entire spectrogram. \cref{fig:loss}(c) demonstrates a very small MSE loss on the validation set when the model converges, suggesting a good reconstruction ability. To show it more straightforward, some examples are visualized in \cref{fig:mae:breth},  \cref{fig:mae:cough}, \cref{fig:mae:lung}. From the visualization, it is clear that our pretrained encoder can capture both the local and global distribution of the spectrograms and the decoder can accurately recover the original information. 

\newpage

\subsection{Additional Evaluation Results}
\label{app:eval}
\begin{figure*}[th]
    \centering
    \subfigure[Health Condition Inference]{\includegraphics[height=5.5cm]{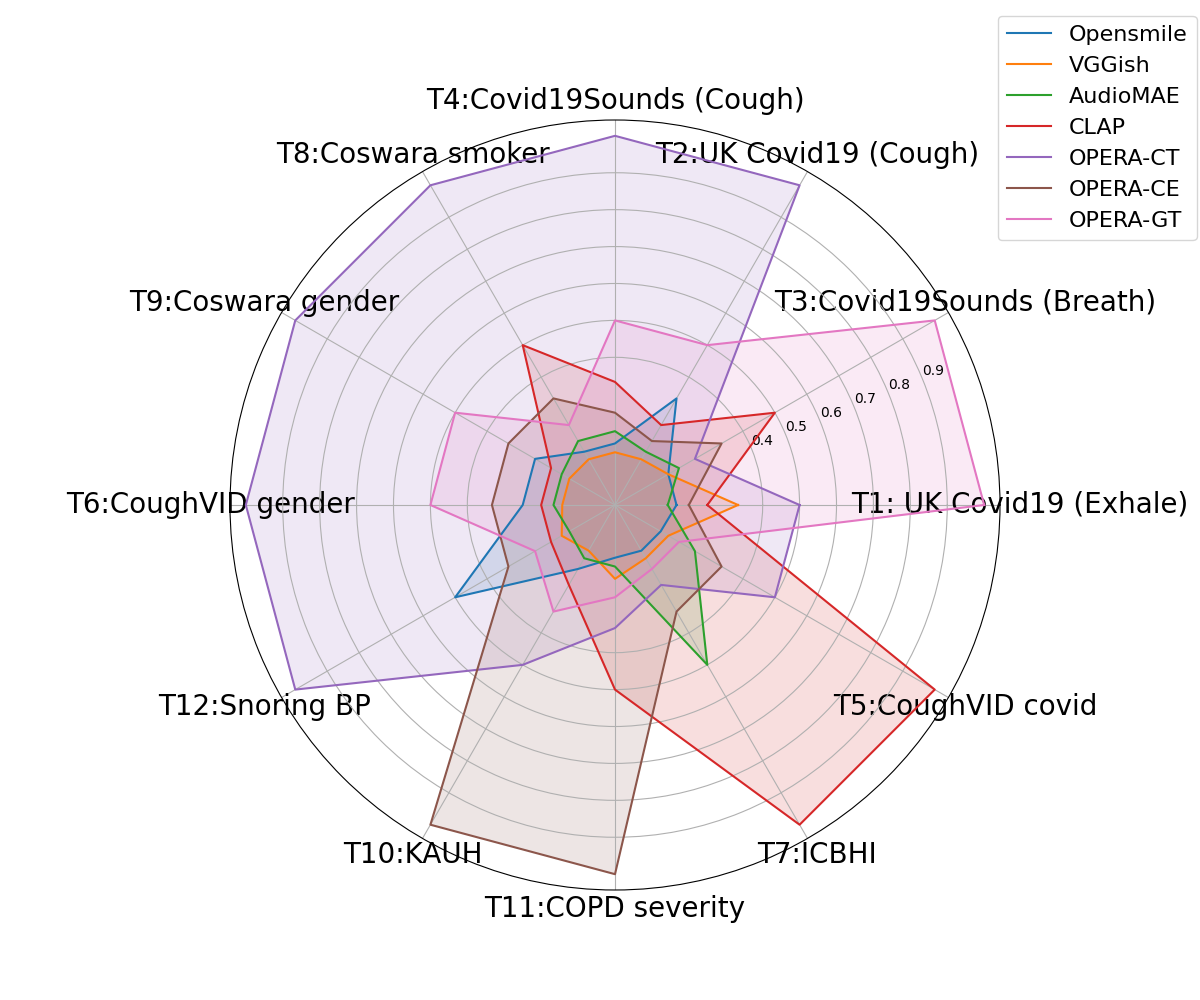} \label{fig:radar:health}}
    \subfigure[Lung Function Estimation]{ \includegraphics[height=5.5cm]{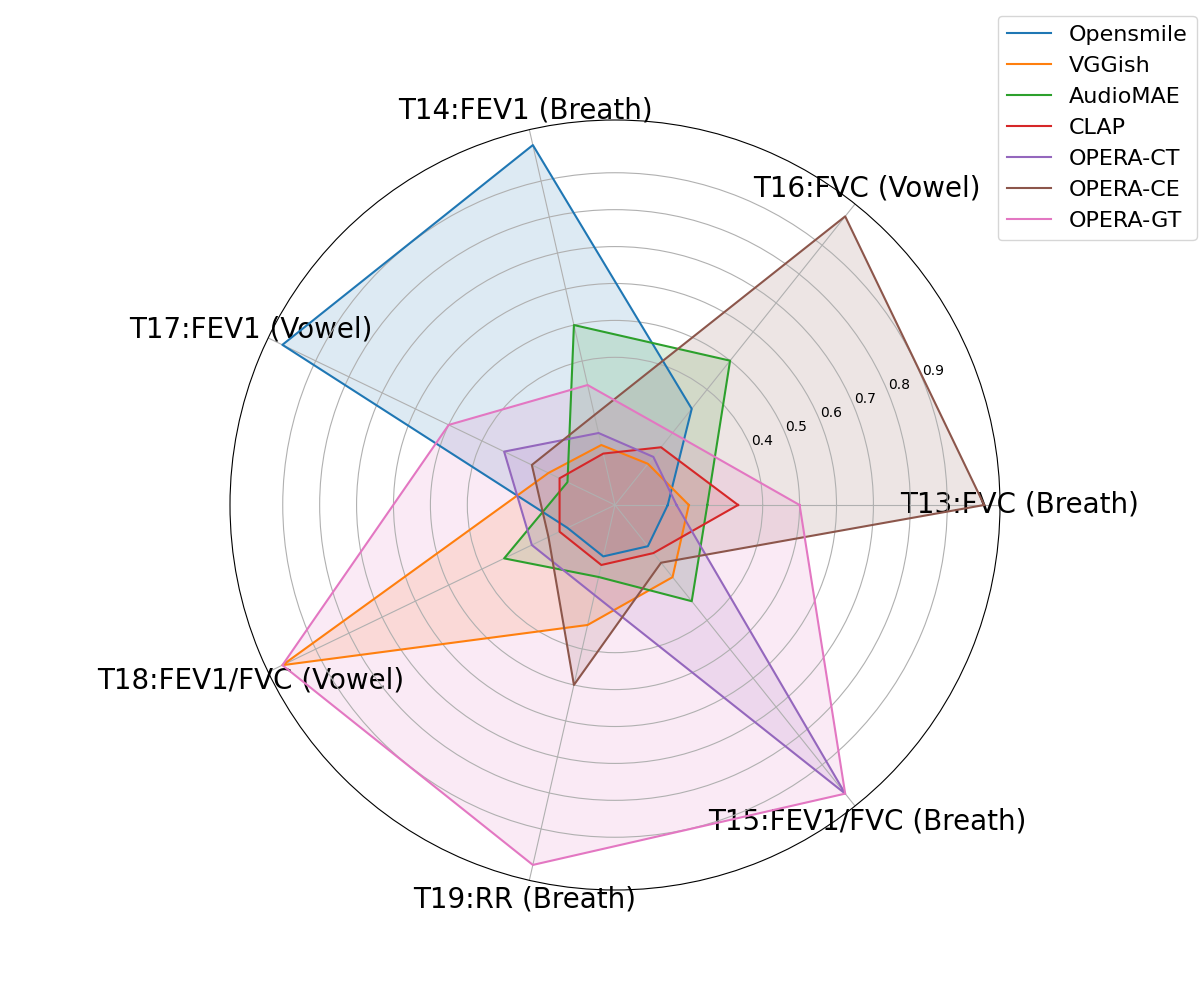} \label{fig:radar:lung}}
     \setlength{\abovecaptionskip}{0.1cm}
    \caption{Radar plot of reciprocal ranks on two groups of tasks. 
    }
    \label{fig:radar}
\end{figure*}

\cref{tab:rank} summarized the over mean reciprocal ranks, with the reciprocal ranks of all the 19 tasks detailed in \cref{fig:radar}.

\subsubsection{Another Metric for Lung Function Estimation Tasks}
While AUROC, used for classification, ranges from 0.5 to 1, MAE, used for regression, doesn't have a bounded range for comparison. Hence, here we additionally report the relative error for the estimation measured by MAPE (Mean Absolute Percentage Error) in \cref{tab:res:lung_mape}. MAPE ranges from 0 to 1, with a lower value indicating better estimations.

\begin{table}[th]
\caption{MAPE on lung function estimation tasks (\textcolor{red}{lower} is better). The best model per task is highlighted. We report mean and standard deviation across subjects. }
\centering
\begin{adjustbox}{width=0.99\textwidth,center} 
\begin{tabular}{ll|c|ccc|cccccc|c}
\toprule
ID & Task Abbr.  & Opensmile     & VGGish        & AudioMAE      & CLAP          & \textbf{OPERA-CT} & \textbf{OPERA-CE} & \textbf{OPERA-GT} &\\
  \midrule

T13 & FVC (Breath)          &0.329 $\pm$ 0.338	&0.298 $\pm$ 0.252	&0.299 $\pm$ 0.245	&0.295 $\pm$ 0.222	&0.304 $\pm$ 0.259	&\cellcolor{mypink}0.278 $\pm$ 0.261	&0.291 $\pm$ 0.247                                                     &\textcolor{red}{\checkmark*}\\
T14 & FEV1 (Breath)         &\cellcolor{mypink}0.353 $\pm$ 0.469	&0.394 $\pm$ 0.444	&0.392 $\pm$ 0.480	&0.396 $\pm$ 0.435	&0.399 $\pm$ 0.449	&0.381 $\pm$ 0.447	&0.392 $\pm$ 0.466                                                     &  \\
T15 & FEV1/FVC (Breath)     &0.178 $\pm$ 0.219	&0.167 $\pm$ 0.165	&0.164 $\pm$ 0.163	&0.174 $\pm$ 0.177	&\cellcolor{mypink}0.161 $\pm$ 0.152	&0.166 $\pm$ 0.149	&0.162 $\pm$ 0.150                                                     &\textcolor{red}{\checkmark*}\\
T16 & FVC (Vowel)           &0.277 $\pm$ 0.238	&0.294 $\pm$ 0.246	&0.280 $\pm$ 0.253	&0.292 $\pm$ 0.247	&0.292 $\pm$ 0.233	&\cellcolor{mypink}0.264 $\pm$ 0.260	&0.293 $\pm$ 0.255                                                     &\textcolor{red}{\checkmark*}\\
T17 & FEV1 (Vowel)          &\cellcolor{mypink}0.342 $\pm$ 0.363	&0.396 $\pm$ 0.446	&0.417 $\pm$ 0.462	&0.402 $\pm$ 0.409	&0.359 $\pm$ 0.372	&0.398 $\pm$ 0.455	&0.368 $\pm$ 0.440                                                      &\textcolor{red}{~~~*}\\    
T18 & FEV1/FVC (Vowel)      &0.175 $\pm$ 0.183	&\cellcolor{mypink}0.167 $\pm$ 0.164	&\cellcolor{mypink}0.167 $\pm$ 0.157	&0.176 $\pm$ 0.170	&\cellcolor{mypink}0.167 $\pm$ 0.153	&0.171 $\pm$ 0.162	&\cellcolor{mypink}0.167 $\pm$ 0.158                                                     & \textcolor{red}{\checkmark*}\\
T19 & Breathing Rate        &0.212 $\pm$ 0.080	&0.205 $\pm$ 0.080	&0.207 $\pm$ 0.086	&0.207 $\pm$ 0.084	&0.207 $\pm$ 0.099	&0.193 $\pm$ 0.065	&\cellcolor{mypink}0.186 $\pm$ 0.071                                                     &\textcolor{red}{\checkmark*}\\

  \bottomrule     
\end{tabular}
\end{adjustbox}
\label{tab:res:lung_mape}
\end{table}

\newpage
\subsubsection{Fine-tuning Performance}

Apart from the standard linear evaluation, we also explore the effect of fine-tuning in improving the performance, using some of the tasks with a comparatively sufficient number of samples.

For OPERA-CE, due to the small number of parameters that could easily overfit and forget the pretraining, we freeze two-thirds of the blocks and only fine-tune the first 5 blocks dealing with the input data (along with the classification head). For all other models and baselines, we fine-tune the entire model together with the classifier.

In addition to the result for Task 4 detailed in \cref{sec:discussion}, the performance of Task 7 and 12 after fine-tuning are presented in \cref{tab:finetune:icbhi} and \cref{tab:finetune:ssbpr}. It is obvious that the performance can be greatly improved after fine-tuning, and the two transformer-based OPERA models demonstrate superior performance.





\begin{table}[th]
\centering
\caption{AUROC (\textcolor{red}{higher} is better) for linear probing and finetuning on T7 (COPD detection). Best model highlighted.}
\label{tab:finetune:icbhi}
\begin{adjustbox}{width=0.99\textwidth,center} %
\begin{tabular}{lccccccccc} 
\toprule
   Method   & \# Train & AudioMAE  & CLAP  & OPERA-CT  & OPERA-CE  & OPERA-GT  \\

 \midrule
\textbf{Linear}             & 828   & 0.886 $\pm$ 0.017 & \cellcolor{mypink}{0.933 $\pm$ 0.005}& 0.855 $\pm$ 0.012                                                           & 0.872 $\pm$ 0.011                                                           & 0.741 $\pm$ 0.011                           \\
\textbf{Fine-tune}           & 828 & 0.984 $\pm$ 0.012	& 0.980 $\pm$ 0.007	& 0.957 $\pm$ 0.024	& 0.808 $\pm$ 0.032	& \cellcolor{mypink}{0.986 $\pm$ 0.006}\\
\bottomrule
\end{tabular}
\end{adjustbox}
\vspace{-5pt}
\end{table}

\begin{table}[th]
\centering
 \caption{AUROC (\textcolor{red}{higher} is better) for linear probing and finetuning on T12 (snoring based body position recognition). Best model highlighted.}
\label{tab:finetune:ssbpr}
\begin{adjustbox}{width=0.99\textwidth,center} %
\begin{tabular}{lcccccccc} 
\toprule
   Method   & \# Train & AudioMAE  & CLAP  & OPERA-CT  & OPERA-CE  & OPERA-GT  \\

 \midrule
\textbf{Linear}             & 7468 & 0.649 $\pm$ 0.001 & 0.702 $\pm$ 0.001                                                           & \cellcolor{mypink}{0.781 $\pm$ 0.000} & 0.769 $\pm$ 0.000                                                           & 0.742 $\pm$ 0.001       \\
    \textbf{Fine-tune}           & 7468                      & 0.981 $\pm$ 0.002	& 0.935 $\pm$ 0.004 &	\cellcolor{mypink}{0.994 $\pm$ 0.001} &	0.981 $\pm$ 0.002 &	0.986 $\pm$ 0.003         \\
\bottomrule
\end{tabular}
\end{adjustbox}
\end{table}

\subsubsection{Cross-domain Zero-shot Performance}
\label{app:zeroshot}

\reb{Zero-shot capacity is an particularly interesting trait for foundation models, especially LLM-based models. Though this is uncommon for models trained solely with unlabeled non-textual data, we also explore cross-domain zero-shot performance following~\cite{jin2023time}. We train a linear probe on source Task A and test it on target Task B, using T6 $\rightarrow$ T9 and T7 $\rightarrow$ T10 as examples, given their similarity (ref. Table 2). Table below shows that OPERA-CT outperforms the baselines.}

\begin{table}[th]
    \centering
    \caption{AUROC (\textcolor{red}{higher} is better) for cross domain zero-shot performance. Best model highlighted.}
    \label{tab:zeroshot}
    \begin{adjustbox}{width=0.99\textwidth,center}
    \begin{tabular}{lccccccc}
    \toprule
   Method   &  Opensmile & VGGish & AudioMAE  & CLAP  & OPERA-CT   \\
\midrule
T6 $\rightarrow$ T9 & 	0.534 $\pm$ 0.048	& 0.537 $\pm$ 0.025	& 0.472 $\pm$ 0.003 & 0.457 $\pm$ 0.005	& \cellcolor{mypink}{}0.600 $\pm$ 0.009 \\
T7 $\rightarrow$ T10 & 	0.682 $\pm$ 0.014	& 0.588 $\pm$ 0.002	& 0.692 $\pm$ 0.003	& 0.722 $\pm$ 0.002	& \cellcolor{mypink}{}0.823 $\pm$ 0.001 \\
\bottomrule
    \end{tabular}
    \end{adjustbox}
\end{table}

\newpage

\subsubsection{Performance for different model architectures}

\reb{To investigate whether models trained using OPERA data consistently outperforms models trained with general audio data, comparison using consistent model architectures is also important. We used the same ViT from AudioMAE in OPERA-GT. Similarly, for CNNs, we pretrained VGG (same as VGGish) and CNN14 (as CLAP) using the contrastive objective. While in the main paper we chose to showcase OPERA-CE for its competitive performance and potential in constrained scenarios, we include the results here.  The better performance of our models suggests the superiority of our curated respiratory audio data and pretrained models for respiratory health. }

\begin{table}[th]
    \centering
    \caption{Average AUROC (\textcolor{red}{higher} is better) for the health condition inference tasks.}
    \label{tab:zeroshot}
    \begin{tabular}{lccccccc}
    \toprule
   Model   &  VGG & CNN14  & ViT   \\
   \midrule
   General audio	& 0.584 &	0.676 &	0.627 \\
   OPERA data	& 0.653	& 0.692	& 0.674 \\
\bottomrule
    \end{tabular}
\end{table}

\subsubsection{Performance for a hybrid model}

Given that contrastive and generative pretraining objectives bring different strengths and weaknesses, we also explored training a model that combines both. Using the ViT encoder, we employed a projection head for contrastive learning and a decoder to reconstruct the spectrogram. Preliminary results indicate that while this combined objective yields a model with more balanced performance, it does not consistently outperform the single-objective pretraining approach. We report the performance in \cref{tab:hybrid} and \cref{tab:hybrid2}, which can be compared with \cref{tab:res:health} and \cref{tab:res:lung}.

\begin{table}[th]
    \centering
    \caption{AUROC (\textcolor{red}{higher} is better) of the hybrid model for the health condition inference tasks.}
    \label{tab:hybrid}
    \begin{adjustbox}{width=0.99\textwidth,center}
    \begin{tabular}{lccccccccccccc}
    \toprule
   \textbf{Task ID}         & T1            & T2            & T3                     & T4                     & T5            & T6            & T7            & T8            & T9            & T10           & T11           & T12           \\
   \midrule
\textbf{Hybrid} & 0.575  & 0.692  & 0.622  & 0.711 & 0.558 & 0.730 & 0.886 & 0.671  & 0.759  & 0.652  & 0.655  & 0.737 \\
\bottomrule
    \end{tabular}
    \end{adjustbox}
\end{table}

\begin{table}[th]
    \centering
    \caption{MAE (\textcolor{red}{lower} is better) of the hybrid model for the health condition inference tasks.}
    \label{tab:hybrid2}
    \begin{adjustbox}{width=0.6\textwidth,center}
    \begin{tabular}{lccccccccccc}
    \toprule
Task ID                  & T13           & T14           & T15           & T16           & T17           & T18           & T19       \\
\midrule
\textbf{Hybrid} & 0.886 & 0.797  & 0.124  & 0.889  & 0.805  & 0.133  & 2.457  \\
\bottomrule
    \end{tabular}
    \end{adjustbox}
\end{table}

\newpage
\subsubsection{Significance tests}

\reb{We conducted significance tests for all tasks and the p values indicating significance is shown in \cref{tab:sigtest}. Compared to the baselines, our models show a significant improvement in most cases. When compared to the best baseline, OPERA-CT performs better (a higher average of AUROC) on 8 tasks, with 5 of these improvements being statistically significant. Our github repo also provides an easy-to-use significance test function for benchmarking purposes and further use.
}

\begin{table}[ht]
\centering
\caption{P-values for significance tests (t-test) for Tasks 1-12. Significant values are highlighted in yellow (p<0.01). The cases where OPERA models outperform the best baseline are underlined.}
\label{tab:sigtest}
\begin{adjustbox}{width=0.8\textwidth,center}
\begin{tabular}{lcccccc}
\toprule
\textbf{Dataset} & \textbf{ID} & \textbf{Best baseline} & \textbf{OPERA-CT} & \textbf{OPERA-CE} & \textbf{OPERA-GT} \\
\midrule
UK COVID-19      & T1          & VGGish                 & \cellcolor{mypink}{}\uline{0.0001}    & \cellcolor{mypink}{}0.0002            & \uline{0.4230}    \\
                 & T2          & CLAP                   & \cellcolor{mypink}{}\uline{0.0000}    & \cellcolor{mypink}{}\uline{0.0022}    & \cellcolor{mypink}{}\uline{0.0000}    \\
COVID-19 Sounds  & T3          & CLAP                   & \cellcolor{mypink}{}0.0075            & \uline{0.2155}    & \uline{0.9161}    \\
                 & T4          & CLAP                   & \uline{0.0558}    & \cellcolor{mypink}{}0.0000            & \cellcolor{mypink}{}0.0011            \\
CoughVID         & T5          & CLAP                   & \cellcolor{mypink}{}0.0003            & \cellcolor{mypink}{}0.0003            & \cellcolor{mypink}{}0.0000            \\
                 & T6          & CLAP                   & \cellcolor{mypink}{}\uline{0.0000}    & \cellcolor{mypink}{}\uline{0.0000}    & \cellcolor{mypink}{}\uline{0.0000}    \\
ICBHI            & T7          & CLAP                   & \cellcolor{mypink}{}0.0000            & \cellcolor{mypink}{}0.0000            & \cellcolor{mypink}{}0.0000            \\
Coswara          & T8          & CLAP                   & \uline{0.1586}    & 0.8547            & \cellcolor{mypink}{}0.0000            \\
                 & T9          & Opensmile              & \cellcolor{mypink}{}\uline{0.0000}    & \cellcolor{mypink}{}\uline{0.0000}    & \cellcolor{mypink}{}\uline{0.0000}    \\
KAUH             & T10         & CLAP                   & \uline{0.0183}    & \cellcolor{mypink}{}\uline{0.0003}    & \uline{0.9875}    \\
Respiratory@TR   & T11         & CLAP                   & 0.7182            & \uline{0.0439}    & 0.4200            \\
SSBPR            & T12         & Opensmile              & \cellcolor{mypink}{}\uline{0.0027}    & 0.9944            & \cellcolor{mypink}{}0.0000        \\   
\bottomrule
\end{tabular}
\end{adjustbox}
\end{table}

\end{document}